\documentclass[lettersize,journal]{IEEEtran}

\usepackage{amsmath,amsfonts}
\usepackage{algorithmic}
\usepackage{algorithm}
\usepackage{array}
\usepackage{textcomp}
\usepackage{stfloats}
\usepackage{url}
\usepackage{verbatim}
\usepackage{graphicx}
\usepackage{xcolor}
\usepackage{cite}
\usepackage{booktabs}
\usepackage{subfigure}
\usepackage{multirow}
\usepackage{makecell}

\begin{document}


\title{Adaptive Model-Based Transfer Learning \\for Dynamic HVAC Control}


\author{Quang-Thang Le,
    Kevin Wijaya,
    Hsin-Yi Lai,
    Che-Kai Liu,
    Ching-Chun Huang
\thanks{\(^{1}\)Department of Computer Science, National Yang Ming Chiao Tung University, Taiwan
{\tt\small lqthang2000.ee12@nycu.edu.tw}}%
\thanks{\(^{2}\)Department of Computer Science, National Yang Ming Chiao Tung University, Taiwan
{\tt\small kevinwijaya.ee13@nycu.edu.tw}}%
\thanks{\(^{3}\)Department of Computer Science, National Yang Ming Chiao Tung University, Taiwan
{\tt\small laisy.ee10@nycu.edu.tw}}%
\thanks{\(^{4}\)Department of Computer Science, National Yang Ming Chiao Tung University, Taiwan
{\tt\small ckliu.cs11@nycu.edu.tw}}%
\thanks{\(^{5}\)Department of Computer Science, National Yang Ming Chiao Tung University, Taiwan
{\tt\small chingchun@nycu.edu.tw} (Corresponding Author)}%
}

\maketitle

\begin{abstract}
In this paper, we aim to automate the adjustment of air handling unit (AHU) setpoints within heating, ventilation, and air conditioning (HVAC) systems to maintain indoor temperatures at user-specified levels. A key challenge lies in obtaining sufficient high-quality sensor data from real buildings. To address this, we explore transfer learning and leverage simulation software to generate training data. We propose an adaptive model-based transfer learning approach for dynamic HVAC control, where the agent directly controls the source domain under conditions identical to the target domain. This eliminates the need for extensive target-specific knowledge to define data generation schedules and reduces the risk of collecting irrelevant samples, while also providing greater flexibility during learning. At the control level, we enhance performance through physics rule embedding, which ensures physical consistency, and long-term-aware setpoint selection strategy, which mitigates abrupt setpoint changes. Finally, to accelerate and stabilize deployment in new buildings, we enable knowledge transfer directly between similar real-world buildings, reducing the need to construct virtual source domains repeatedly.
\end{abstract}



\section{Introduction}
\IEEEPARstart {T}{he} control of heating, ventilation, and air conditioning (HVAC) is essential for effective building management, given that HVAC systems typically represent a significant portion of overall energy consumption. Traditional methods like Proportional-Integral-Derivative (PID) control are widely used due to their simplicity, but tuning and achieving reliable performance remain challenging \cite{ref24}. Deep Reinforcement Learning (DRL) allows agents to learn and adapt continuously; however, it also suffers from challenges in real-world exploration due to high costs and long convergence times \cite{ref9, ref10, ref23}. Advanced techniques such as Model Predictive Control (MPC) provide promising control capabilities. Meanwhile, emerging methods like neural networks (NN) are also demonstrating potential. For example, \cite{ref3} developed a predictive model for an air handling unit using NN and successfully applied it to control a large central air conditioning system. However, the implementation of data-driven models still faces obstacles, especially in acquiring sufficient historical data for training. This problem becomes more complex due to the dynamic nature of building environments, which requires considerable time and effort to ensure equipment functionality and collect high-quality data. Therefore, applying these methods in real-world scenarios remains difficult. Moreover, settings in actual buildings tend to remain monotonous, which introduces bias in the collected data.

The most straightforward solution to address the scarcity of high-quality, unbiased data is to operate real-world HVAC systems under varying control strategies, rather than relying on fixed schedules that may introduce bias. However, this may lead to occupant discomfort, which is not acceptable in practice. OSLN \cite{ref1} attempted to mitigate data bias by using biased real-world data combined with a simple temperature variation principle to generate more diverse synthetic data. However, the accuracy of the synthetic data was questionable. Additionally, the setpoint variations, which led to unstable indoor conditions and increased wear on HVAC equipment.

{Many prior studies rely on static pretraining using synthetic datasets generated from simulation. For example, \cite{ref19} used EnergyPlus to produce a large dataset for pretraining, followed by fine-tuning with real sensor data. However, static datasets often contain redundant or biased samples and are based on predefined schedules, which prevent the agent from experiencing a wide variety of control situations. This limitation reduces the model's ability to learn robust strategies and generalize to real building operations, since pretraining does not expose the agent to rare or extreme conditions.

Recent work by NVIDIA on Cosmos AI \cite{nvidia-cosmos-2025, ren-cosmos-drive-dreams-2025, nvidia-cosmos-transfer1-2025} highlights a growing trend in leveraging digital twins, which consist of high-fidelity simulation environments and virtual replicas of physical systems, combined with synthetic data generation to safely train and evaluate physical AI agents in simulation prior to real-world deployment. This approach helps to mitigate risks and reduce costs by enabling extensive testing and learning in a controlled and repeatable setting. Inspired by this direction, our work applies a similar concept to the domain of HVAC control by utilizing a physics-based simulator that accurately models building thermal dynamics as a digital twin. This allows us to create a realistic and interactive environment where AI control strategies can be tested and refined before being applied in the real world.

Unlike static pretraining, we introduce dynamic pretraining in which the model actively interacts with a simulated environment built in EnergyPlus. Instead of relying on fixed setpoint schedules, the model adjusts control setpoints and observes the resulting temporal room temperature responses. Through dynamic control, the agent explores diverse scenarios, including rare or extreme conditions that would be unsafe or uncomfortable to test in real buildings. This allows the model to capture variations that static datasets cannot provide. The pretrained model develops a richer understanding before adapting to real buildings.

Building on this foundation, we propose an adaptive model-based transfer learning framework that integrates dynamic pretraining, transfer, and online adaptation. The model learns a neural network–based dynamic representation of room temperature responses in a simulated source domain using EnergyPlus. The pretrained model is then transferred to the target building and fine-tuned with a small amount of real sensor data. Leveraging dynamic pretraining, the model requires only slight adaptation to the target building and continues to adapt through online learning. This allows the model to adapt with target domain condition and reducing the need for manual setpoint design.}

Based on promising virtual-to-real transfer results, our HVAC control agent can be extended to more buildings. For buildings sharing similar characteristics such as layout and size, existing stable buildings can serve as source domains instead of creating new virtual models. This approach reduces setup time and facilitates faster, more reliable expansion of HVAC control systems.

{Furthermore, to ensure the model produces physically consistent predictions, we embed known physical laws into the training process using constraint-based loss functions. Even with dynamic pretraining and agent-controlled operation in real buildings, the model may still generate outputs that violate fundamental physical relationships. This embedding of physical rules ensures that the model respects essential input-output relationships and expected state transitions. By enforcing these rules, the model generates reliable and physically consistent predictions.}

{
Additionally, frequent abrupt setpoint changes in HVAC systems often cause temperature oscillations, increased energy waste, and reduced occupant comfort. To address this, we introduce a long-term-aware setpoint selection strategy that predicts future room temperature and evaluates candidate setpoints based on both immediate alignment with the target and long-term stability. By considering the predicted final stable temperature, the strategy enables smoother setpoint adjustments, reducing fluctuations while balancing comfort and energy efficiency over time.
}

In summary, we address the potential limitations of existing HVAC control methods \cite{ref1} by proposing an adaptive model-based transfer learning approach to dynamically pretrain the model, with the aim of overcoming challenges such as the difficulty of collecting high-quality training data. We briefly outline our main contributions as follows:
\begin{itemize}
    \item To address challenges in data collection from real buildings and facilitate stable deployment in new buildings, we propose an adaptive model-based transfer learning method. This approach involves creating a virtual building resembling the target real building and employing dynamic control to adapt the agent to the virtual environment. Subsequently, we transfer the agent, familiar with the virtual environment's thermal dynamics, for application in real buildings, effectively bridging the gap between virtual and real domains.
    \item We introduce a physics rule embedding method that integrates known input-output relationships and expected state transitions as constraints during training, combined with a long-term-aware control strategy. This approach ensures the model respects physical laws and {considers the future effects of each setpoint}, resulting in smoother and more physically consistent HVAC control.
    \item To accelerate the expansion of the HVAC control system, we implement a configuration transfer mechanism between real-world buildings. By leveraging the operational HVAC control systems of similar existing buildings, we transfer their stable control setups to new target buildings, thereby speeding up installation and enhancing system scalability.
\end{itemize}

\section{Related Work}
\subsection{Approaches for HVAC Control}

Controlling HVAC systems is a critical aspect of building management, prompting the development of various approaches. For instance, on/off control offers a straightforward solution where the cooling unit activates when system temperature exceeds a threshold and deactivates below it. However, this method lacks adaptability and may lead to temperature overshoots and energy inefficiencies \cite{ref8}. To address these shortcomings, PID control was introduced, which continuously adjusts controller parameters based on the difference between feedback and setpoint signals \cite{ref24}. Despite its effectiveness in managing overshoots, PID tuning can be time-consuming and challenging for complex systems, especially those with multiple inputs and outputs (MIMO).

Advanced control methods like Model Predictive Control (MPC) aim for energy efficiency and effectiveness in MIMO systems. For instance, \cite{ref5} and \cite{ref4} segmented cooling systems into subsystems, employing dynamic models to design controllers. These models, built on physical principles, offer a strong theoretical foundation but entail complex modeling and specific conditions. In contrast, data-driven approaches, such as those using Deep Neural Networks (DNN) and recurrent neural networks (RNN) \cite{ref3, ref2}, have gained traction due to their flexibility and lower complexity. By mapping extensive data, these methods capture system dynamics effectively, enhancing stability during initial operation. However, they heavily rely on high-quality training data and precise algorithms.

Some studies explore control methods based on Deep Reinforcement Learning (DRL), like those in \cite{ref9, ref10, ref23}. These methods formulate control as a Markov Decision Process (MDP), allowing agents to learn and adapt continuously. Despite their ability to handle uncertainties and optimize energy costs, DRL approaches face challenges in exploration within real environments due to cost and convergence time \cite{ref11}.

Moreover, hybrid approaches are emerging, such as the Online Self-Learning Network (OSLN) proposed in \cite{ref1}. OSLN combines the stability benefits of advanced control methods with the self-learning advantages of DRL. In the training stage, synthetic data is generated to overcome bias, while an optimal setpoint selection strategy, based on dynamic model predictions, guides operation. However, this strategy's pursuit of minimal temperature differences may lead to drastic setpoint changes.

\subsection{Transfer Learning}

Transfer learning has gained widespread use across various fields to mitigate the scarcity of training data. It can be broadly categorized into three main types: instance-based TL, feature-based TL, and model-based TL. Among these, instance-based TL is the most straightforward and intuitive to implement. This approach involves selecting samples from the source domain that are most relevant to the target domain and assigning appropriate weights based on their relevance. For instance, \cite{ref15} incorporated auxiliary training sets related to the target problem into the training process of the target forecasting model, while \cite{ref16} extended target data by transferring data from related cities. However, instance-based TL heavily relies on a strong correlation between source and target data, which can be challenging to achieve in practice and may lead to negative transfer when this correlation is absent.

In contrast, feature-based TL can effectively address significant domain gaps between source and target domains. It typically involves mapping both domains' feature spaces to a common feature space, facilitating the analogization of the target task to the source task and overcoming the limitation of training data in the target domain. For instance, \cite{ref12} aligned domain-invariant features of the two domains and trained domain-specific features simultaneously, while \cite{ref13} automatically mapped sensor and control points of a new building to common representations without manual intervention. Although feature-based TL excels at extracting or sharing static features, it may fall short in capturing dynamic changes in the environment, which is crucial for HVAC control.

{According to statistics from \cite{ref6}, a majority of HVAC control methods rely on model-based TL.} Therefore, model-based TL emerges as a more suitable choice, typically involving the transfer of model parameters trained in the source domain to the target domain. For example, \cite{ref7} fixed most layers of a model pre-trained on the source room and fine-tuned the trainable parameters with data from the target room. {However, the controller cannot learn new control strategies on its own. } Similarly, \cite{ref14} fine-tuned layers copied from the source domain with the target dataset. Nevertheless, the approach remains offline and does not support adaptive control.

{
Another example, \cite{ref18} tackles the challenges of long training time and limited scalability in DRL-based HVAC controllers. To address this, the study decomposes the controller into a building-agnostic front-end and a building-specific back-end, enabling knowledge transfer and reducing DRL training time. \cite{ref19} pre-trained a model on a long-term simulated dataset generated by EnergyPlus and fine-tuned it with short-term real measurements from an actual building to improve prediction accuracy. Similarly, \cite{ref20} employed a pre-trained LSTM sequence-to-sequence model from a source building and fine-tuned it for a target building. While these approaches improve training efficiency or predictive performance, the controllers remain fixed and cannot adapt online or dynamically update control strategies on its own, which limits their applicability in dynamic environments.
}

{Overall, while existing model-based transfer learning methods improve prediction or control in specific buildings, they mostly rely on static simulation datasets with fixed setpoint schedules, which may not capture extreme conditions. These limitations indicate that existing methods cannot fully capture dynamic and rare control scenarios, motivating the need for a framework that can safely explore and adapt to diverse strategies. In our framework, the model is dynamically pretrained in a simulated source domain using EnergyPlus and transferred to the target building. It is fine-tuned with a small amount of real sensor data and continues to adapt online, enabling real-time adjustment and reducing the need for manual setpoint design.}

\section{Problem Statement}

\begin{figure}[!t]
    \centering{\includegraphics[width=\columnwidth]{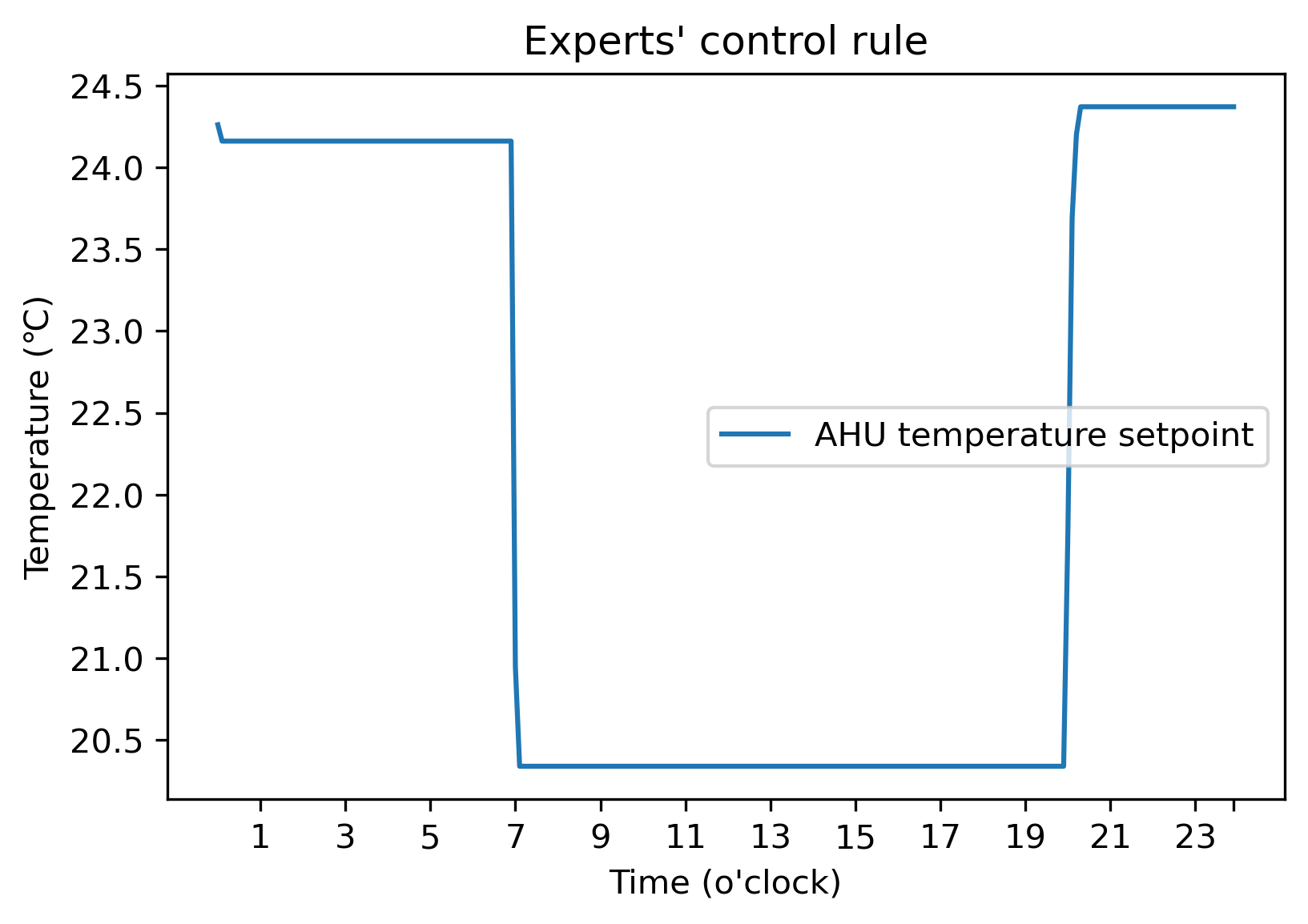}}
    \caption{AHU temperature setting schedule designed by experts. Other than adjusting the settings to a lower level when approaching work and to a higher level when approaching the end of work, there won't be much change in the settings.}
    \label{figure:expert_control_rule}
\end{figure}

\begin{figure}[!t]
    \centering
    \resizebox{\columnwidth}{!}{%
        \includegraphics[width=\columnwidth]{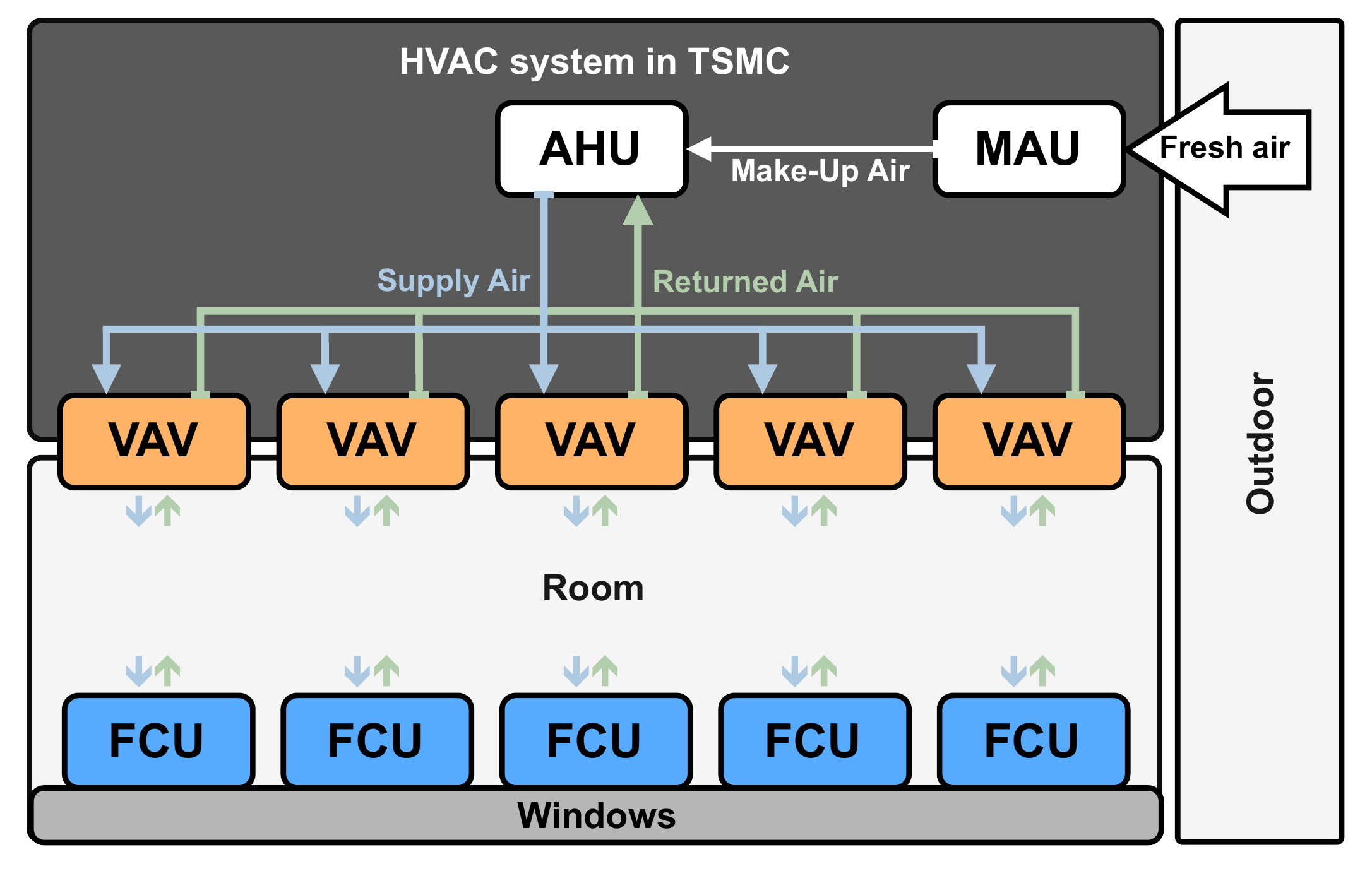}
    }
    \caption{The HVAC system schematic.}
    \label{figure:HVAC}
\end{figure}

As depicted in Fig.~\ref{figure:HVAC}, the HVAC system in our buildings comprises three primary components: the Make-up Air Unit (MAU), Air Handling Unit (AHU), and Variable Air Volume (VAV) system. The MAU extracts fresh outdoor air, which undergoes filtration and cooling processes before being supplied indoors. The AHU recirculates indoor air, controlling supply air temperature and volume to maintain indoor temperature levels. The VAV system regulates supply air volume through adjustable air valves, acting as air outlets. Additionally, large buildings may incorporate independent cooling units called Fan Coil Units (FCU) for further indoor temperature control. The temperature readings at each VAV represent the terminal temperatures in adjacent zones. Our objective is to manage the AHU's supply air temperature setpoint to ensure that {room} temperatures, reflecting indoor temperatures, align with user preferences.

Our objective is to empower the HVAC control agent to swiftly and reliably operate in new buildings, even in scenarios where data collection presents challenges. To achieve this, we divide our framework into two components: the proposed HVAC control method and the proposed transfer learning method.

\section{Proposed Method}
Our framework is structured into two main components: the operational phase of the agent and the transfer learning aspect. Initially, we introduce the construction of the dynamic model and the adopted control strategy during the agent's operation. Following this, we outline the establishment of our source domain, providing insights into the comprehensive process of our transfer learning approach. Further elaboration on these aspects will be provided in subsequent sections.

\subsection{Training Objectives}
Overall, our goal is to develop a dynamic model capable of accurately forecasting future {room} temperatures. The input comprises three components: $S_t$, $E_t$, and $A_t$. Specifically, $S_t$ represents the current states, encompassing the current VAV{/room} temperatures. {$E_t$ denotes uncontrollable factors influencing room temperature that cannot be adjusted. In this work, $E_t$ includes FCU temperatures, VAV and FCU on off status, power supplied, MAU outlet temperature, outdoor temperature, and weather conditions as provided in {Table~\ref{tab:input_description}}. These factors are directly measured by sensors installed in the building and provide real time information about environmental and operational conditions. By defining $E_t$ based on sensor measured variables, the proposed control strategy can adapt to different buildings and operating conditions using real time data, enabling wider applicability across HVAC systems with varying layouts and configurations}. {$A_t$ represents the current AHU setpoint, corresponding to the action selected by the agent to maintain room temperature.}

The output, $S^\prime_{t+\Delta}$, predicts the {room} temperatures at time $t+\Delta$. {In our experiments, $\Delta$ is set to 18 minutes. The 18-minute control cycle was selected based on experiments balancing stability and accuracy. Details of the experimental comparison and justification are provided in Section~\ref{subsubsec:control_cycle_duration}}. Additionally, the output $D^\prime_{t+\Delta}$ estimates the difference between the long-term temperature $S_{long}$, representing the long-term prediction of temperature over time given the setpoint $A_t$, and the short-term temperature $S_{t+\Delta}$. This measure is designed to gauge setpoint stability, offering an approximate estimation of subsequent indoor temperature changes if the setpoint $A_t$ remains unchanged at time $t+\Delta$. In essence, a smaller value of $D^\prime_{t+\Delta}$ associated with a given setpoint indicates greater stability in the environmental conditions.

\begin{figure}[!t]
    \centering{\includegraphics[width=\columnwidth]{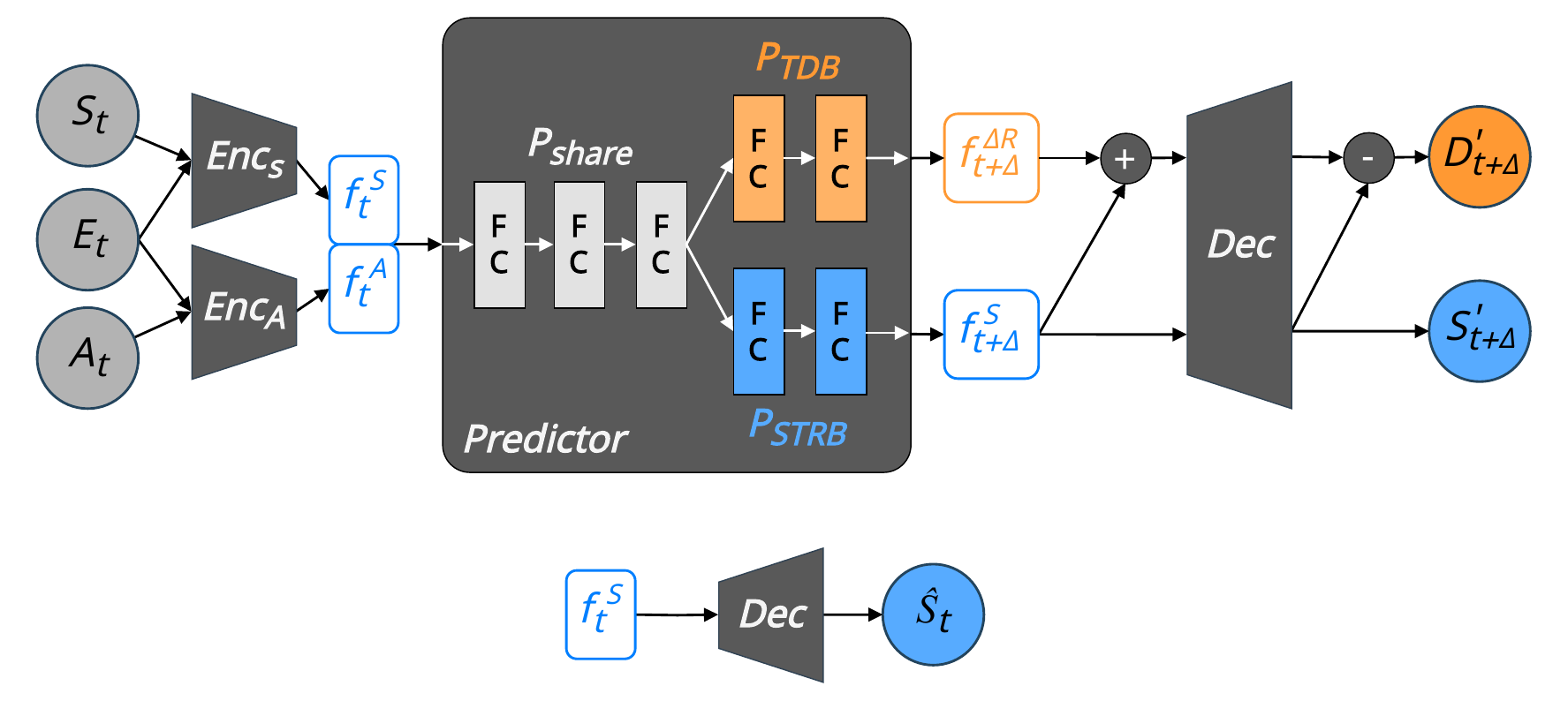}}
    \caption{The architecture of the dynamic model.}
    \label{figure:ED_model}
\end{figure}

{
Before entering the dynamic model, the input features $S_t$, $E_t$, and $A_t$ are normalized using either a standardized scaler or a maximum scaler depending on the type of feature. Room temperatures and other continuous sensor measurements are standardized, while equipment status signals are maximum-normalized. The normalized features are then fed into the state encoder $Enc_S$ and the action encoder $Enc_A$ to produce latent representations for the predictor and decoder.

The proposed dynamic model, illustrated in Fig.~\ref{figure:ED_model}, consists of three main components: encoders, a predictor, and a decoder. The inputs are first processed by the state encoder $Enc_S$ and the action encoder $Enc_A$, producing latent features $f_t^S$ and $f_t^A$, respectively. The state encoder captures the current room temperature $S_t$ and uncontrollable environmental factors $E_t$. The action encoder captures how the AHU setpoint $A_t$ interacts with the environment, producing a latent feature that characterizes the effect of the control action. These encoder outputs are formally defined as:
\begin{equation}
    \label{latent_features}
    \begin{aligned}
    f^S_t &= Enc_S(S_t, E_t) \\
    f^A_t &= Enc_A(E_t, A_t)
    \end{aligned}
\end{equation}

The latent features $f_t^S$ and $f_t^A$ are then concatenated and passed into the predictor, which consists of a shared network $P_{share}$ and two specialized branches. The shared network captures general building dynamics from $S_t$, $E_t$, and $A_t$, producing a joint representation. The temperature difference branch (TDB) $P_{TDB}$ encodes the long-term residual change in temperature $f_{t+\Delta}^{\Delta R}$, which helps maintain setpoint stability by preventing large fluctuations. The short-term regression branch (STRB) $P_{STRB}$ predicts the short-term room temperature $f_{t+\Delta}^S$ for immediate control. The long-term latent representation $f_{t+\Delta}^R$ is obtained by adding the residual feature $f_{t+\Delta}^{\Delta R}$ to the short-term feature $f_{t+\Delta}^S$. Formally, the latent features are defined as:
\begin{equation}
    \label{latent_features}
    \begin{aligned}
    f^{S}_{t+\Delta} &= P_{STRB}(P_{share}([f^S_t, f^A_t]) \\
    f^{\Delta R}_{t+\Delta} &= P_{TDB}(P_{share}([f^S_t, f^A_t])
    \end{aligned}
\end{equation}
}

The decoder produces the predicted short-term temperature $S'_{t+\Delta}$ from the latent feature $f_{t+\Delta}^S$. The difference between the long-term temperature and the short-term temperature, denoted as $D'_{t+\Delta}$, is obtained by subtracting the predicted short-term temperature $S'_{t+\Delta}$ from the decoded long-term latent representation $\mathrm{Dec}(f_{t+\Delta}^R)$. Additionally, the feature $f_t^S$ is directly passed through the decoder without the predictor block to produce the reconstruction output $\hat{S}_t$. The processes of obtaining $D'_{t+\Delta}$, $S'_{t+\Delta}$, and $\hat{S}_t$ are respectively denoted by the functions $F_{TDB}$, $F_{STRB}$, and $F_{REC}$, and can be expressed as follows:
\begin{equation}
    \label{output_S_plus}
    \begin{aligned}
        S^\prime_{t+\Delta} &= F_{STRB}(S_t, E_t, A_t) \\
        &= Dec\bigl(f_{t+\Delta}^S\bigr)
    \end{aligned}
\end{equation}

\begin{equation}
    \label{output_D}
    \begin{aligned}
        D^\prime_{t+\Delta} &= F_{TDB}(S_t, E_t, A_t) \\
        &= Dec\bigl(f_{t+\Delta}^{\Delta R} + f_{t+\Delta}^S\bigr) - S^\prime_{t+\Delta}
    \end{aligned}
\end{equation}

{\begin{equation}
    \label{reconstruction_S}
    \begin{aligned}
        \hat{S}_t &= F_{REC}(S_t, E_t) \\ 
        &= Dec\bigl(f^S_t) 
    \end{aligned}
\end{equation}}

$F_{STRB}$ is optimized by the normal regression loss, as shown in \eqref{regression_loss}. $S_{t+\Delta}$ represents the ground truth and $S'_{t+\Delta}$ represents the prediction of the model. {${Loss}_{{STRB}}$ encourages accurate next-step prediction by minimizing the mean squared error between the predicted and ground-truth states.}
\begin{equation}
    \label{regression_loss}
    Loss_{STRB} = {\lVert S_{t+\Delta} - S'_{t+\Delta} \rVert}^2
\end{equation}

To mitigate frequent and drastic changes in the setpoint, our approach prioritizes long-term temperature stability over short-term fluctuations. This strategic shift prevents {a} blind pursuit of minimal errors, ensuring that stability is not overlooked. Initially, we define $S_{long}$ as the average indoor temperature influenced by setpoint $A_t$ over an extended period. Mathematically, $S_{long}$ is computed as the weighted average temperature indoors, incorporating a decay coefficient $\gamma$ to emphasize temperatures closer in time:
\begin{equation}
    \label{LT_function}
    \begin{aligned}
    S_{long} &= \frac{\sum_{k=t+\Delta}^{\infty} \gamma^{k - t - \Delta} S_k}{\frac{1}{1 - \gamma}} \\
    S_{long} &= (1 - \gamma) \sum_{k=t+\Delta}^{\infty} \gamma^{k - t - \Delta} S_k
    \end{aligned}
\end{equation}

Here, a larger $\gamma$ underscores the significance of temperatures in closer proximity to the present. This definition of $S_{long}$ enables a holistic understanding of the indoor thermal environment's long-term behavior under the influence of $A_t$. {Notably, this quantity cannot be directly measured during training because the future ground-truth temperatures $S_k$ for $k > t+\Delta$ are not available. To address this, we derive a practical approximation for $S_{\text{long}}$ using a recursive formulation, denoted as $Y$, which is then used to fit the model’s prediction $D'_{t+\Delta}$.}

Meanwhile, we utilize equation \eqref{LT_function} to derive the loss function for fitting $D_{t+\Delta}$. {Let $LT^*(S_t, E_t \mid A_t)$ represent the long-term temperature under setpoint $A_t$, which can be written as:
\begin{equation}
    \label{LT_function2}
    \begin{aligned}
    LT^*(S_t, &E_t | A_t) = S_{\text{long}} \\
    LT^*(S_t, &E_t \left| A_t \right) = (1-\gamma)(\sum_{k=t+\Delta}^{\infty}{\gamma^{k-t-\Delta} S_k}) \\
    = (1-\gamma)&(S_{t+\Delta}+\gamma S_{t+2\Delta}+\gamma^2 S_{t+3\Delta}+...+r^\infty S_\infty) \\
    = (1-\gamma)&(S_{t+\Delta}+\gamma (S_{t+2\Delta}+\gamma S_{t+3\Delta}+...+r^\infty S_\infty)) \\
    = (1-\gamma)&(S_{t+\Delta}+\gamma\times \frac{LT^*(S_{t+\Delta}, E_{t+\Delta} \left| A_t \right)}{1-\gamma}) \\
    = (1-\gamma)&\times S_{t+\Delta} + \gamma\times LT^*(S_{t+\Delta}, E_{t+\Delta} | A_t)
    \end{aligned}
\end{equation}}
The equation \eqref{LT_function2} can be rearranged as:
\begin{equation}
    \label{LT_function3}
    \begin{aligned}
        &LT^*(S_t, E_t \left| A_t \right)- \\
        &(S_{t+\Delta}(1-\gamma)+\gamma\times LT^*(S_{t+\Delta}, E_{t+\Delta} \left| A_t \right))=0
    \end{aligned}
\end{equation}
Since we utilize $F_{TDB}(S_t, E_t, A_t)$ to estimate $D_{t+\Delta}$, if the model converges, the following equation must hold:
\begin{equation}
    \label{LT_function4}
    \begin{aligned}
        F_{TDB}(S_t, E_t, A_t)&=S_{long}-S_{t+\Delta} \\
        &=LT^*(S_t, E_t \left| A_t \right)-S_{t+\Delta} \\
        LT^*(S_t, E_t \left| A_t \right)&=F_{TDB}(S_t, E_t, A_t)+S_{t+\Delta}
    \end{aligned}
\end{equation}
Substituting equation \eqref{LT_function4} into equation \eqref{LT_function3}, we obtain:
{\begin{equation}
    \label{LT_function5}
    \begin{aligned}
        &F_{TDB}(S_t, E_t, A_t) + S_{t+\Delta} \\
        &- \Big((1-\gamma) S_{t+\Delta} + \gamma \, LT^*(S_{t+\Delta}, E_{t+\Delta}| A_t) \Big) = 0 \\
        &F_{TDB}(S_t, E_t, A_t) + S_{t+\Delta} \\
        &- S_{t+\Delta} + \gamma S_{t+\Delta} - \gamma LT^*(S_{t+\Delta}, E_{t+\Delta} \mid A_t) = 0 \\
        &F_{TDB}(S_t, E_t, A_t) - \gamma LT^*(S_{t+\Delta}, E_{t+\Delta} \mid A_t) + \gamma S_{t+\Delta} = 0 \\
        &F_{TDB}(S_t, E_t, A_t) - \gamma \big( LT^*(S_{t+\Delta}, E_{t+\Delta} \mid A_t) - S_{t+\Delta} \big) = 0
    \end{aligned}
\end{equation}}
Rearranging equation \eqref{LT_function5}, we obtain:
\begin{equation}
    \begin{aligned}
        &F_{TDB}(S_t, E_t, A_t) \\
        &-\gamma(F_{TDB}(S_{t+\Delta}, E_{t+\Delta}, A_t)+S_{t+2\Delta}-S_{t+\Delta})=0
    \end{aligned}
\end{equation}
{Since the future temperature $S_{t+2\Delta}$ is unknown during training, we replace it with the short-term model prediction $F_{STRB}(S_{t+\Delta}, E_{t+\Delta}, A_t)$:}
{\begin{equation}
    \label{LT_function6}
    \begin{aligned}
        &F_{TDB}(S_t, E_t, A_t)-\gamma\times(F_{TDB}(S_{t+\Delta}, E_{t+\Delta}, A_t) \\
        &+F_{STRB}(S_{t+\Delta}, E_{t+\Delta}, A_t)-S_{t+\Delta})=0
    \end{aligned}
\end{equation}}
Hence, we derive a fully computable target denoted as \(Y\): 
\begin{equation}
    \label{difference_loss_Y}
    \begin{aligned}
        &Y = \gamma\times(F_{TDB}(S_{t+\Delta}, E_{t+\Delta}, A_t) \\
        &+ F_{STRB}(S_{t+\Delta}, E_{t+\Delta}, A_t) - S_{t+\Delta})
    \end{aligned}
\end{equation}
Where $\gamma$ is a weighting factor that captures the long-term effect in the proposed approximation. This formulation is inspired by temporal difference learning, where predictions are updated through bootstrapping using future estimated states. To enhance training stability, a separate target network is employed to estimate $Y$. Finally, the loss function for the long-term difference branch is defined as:  
\begin{equation}
    \label{difference_loss}
    Loss_{TDB} = {\lVert D'_{t+\Delta}-Y \rVert}^2
\end{equation}
{${Loss}_{{TDB}}$ promotes long-term stability by constraining the predicted temporal difference to be consistent with the recursively constructed target $Y$}.

{Moreover, physics rule embedding ensures the model’s predictions remain physically plausible. By incorporating fundamental laws through constraint-based losses, the model respects essential input-output relationships and expected state transitions, producing reliable and consistent predictions even under dynamic, agent-controlled conditions.

In this study, physics rule embedding integrates practical physical constraints of the HVAC system directly into the model’s training objective, rather than using full physical equations.} The first physics-based constraints is the gradient constraint loss function $Loss_{GC}$, which ensures that the model adheres to fundamental physical laws during prediction. This loss describes how each input, such as the current VAV temperature $S_t$, uncontrollable environmental factors $E_t$, and AHU setpoints $A_t$, affects the future VAV temperature $S'_{t+\Delta}$. These rules are incorporated as constraint terms in the loss function and guide the model to produce predictions that obey the expected physical behavior. By imposing this constraint, we mitigate potential biases in the data and maintain the robustness of the model for subsequent operational phases.
\begin{equation}
    \label{derivative_function}
    F = \sum_{\forall element \in S'_{t+\Delta}}{\frac{\partial S'_{t+\Delta}}{\partial X_t}}
\end{equation}

$X_t$ represents the vector obtained by concatenating the inputs $S_t$, $E_t$, and $A_t$. Each element in $X_t$ is associated with a relation label $L$, indicating whether it is positively or negatively correlated with future room temperatures. Each element in $L$ is defined in Equation~\eqref{gradient_constraint_label}, and the relation of each input can be seen in Table~\ref{tab:relation_labels}.
\begin{equation}
    \label{gradient_constraint_label}
    L = 
    \begin{cases}
        1,  & \text{if positively related} \\
        -1, & \text{if negatively related}
    \end{cases}
\end{equation}
\begin{table}[h]
    \centering
    \caption{Relation Labels for Each Feature}
    \label{tab:relation_labels}
    \resizebox{\columnwidth}{!}{%
    \begin{tabular}{lc}
        \toprule
        \textbf{Feature} & \textbf{Relation Label (L)} \\
        \midrule
        AHU setpoint ($A_t$) & +1 \\
        VAV temperature ($S_t$) & +1 \\
        FCU temperature ($E_t$) & +1 \\
        VAV on/off ($E_t$) & +1 \\
        FCU on/off ($E_t$) & -1 \\
        Power supplied ($E_t$) & +1 \\
        MAU outlet temperature ($E_t$) & +1 \\
        Outdoor temperature ($E_t$) & +1 \\
        Weather condition ($E_t$) & -1 \\
        \bottomrule
    \end{tabular}
    }
\end{table}

After defining the relation labels for each input feature, we compute the gradient constraint loss function in Equation~\eqref{gradient_constraint_loss}, where \(\odot\) denotes element-wise multiplication between the relation labels \(L\) and the gradient matrix \(F\). This loss penalizes the model whenever the gradient violates the expected physical relationship.
\begin{equation}
    \label{gradient_constraint_loss}
    Loss_{GC} = 
    \sum_{\forall element \in X_t} \max(0, - L \odot F)
\end{equation}

The second physics-based constraint is the residual constraint loss $Loss_{RC}$, defined by \eqref{residual_constraint_loss} to encourage the predicted long-term residual $D'_{t+\Delta}$ to follow the same direction as the actual change in state $S_{t+\Delta} - S_t$. This loss penalizes residuals that contradict the observed state transition. By applying this constraint, the model is encouraged to produce residuals aligned with the true state evolution, improving prediction stability and physical consistency.
\begin{equation}
    \label{residual_constraint_loss}
    Loss_{RC} = \max\left(0, - (S_{t+\Delta} - S_t) \odot D'_{t+\Delta}\right)
\end{equation}

Finally, the overall loss for the physics rule embedding is the sum of the two physics-based constraints defined above. This loss will later be integrated into the total training loss.
\begin{equation}
    \label{physics_based_constraints}
    Loss_{\text{physics-based constraints}} = Loss_{GC} + Loss_{RC}
\end{equation}

The reconstruction loss \(Loss_{recons}\) {ensures that the latent representation preserves essential state information by penalizing reconstruction errors}. It is defined as the mean squared error (MSE) between the reconstructed state \(\hat{S}_t\) and the true state \(S_t\):
\begin{equation}
    \label{backcast_loss}
    Loss_{recons} = \left\| \hat{S}_{t} - S_t \right\|^2
\end{equation}

In summary, the objective loss of our dynamic model training is expressed as:
\begin{equation}
    \label{total_loss}
    \begin{aligned}
        Loss_{total} = \ &
        \underbrace{Loss_{STRB}}_{\text{short-term prediction}} +
        \underbrace{Loss_{TDB}}_{\text{long-term stability}} \\
        & +
        \underbrace{Loss_{GC} + Loss_{RC}}_{\text{physics-based constraints}} +
        \underbrace{Loss_{recons}}_{\text{state reconstruction}}
    \end{aligned}
\end{equation}

The optimization of TDB relies on \eqref{difference_loss} to iteratively refine its parameters. However, the reliance on unconverged TDB and STRB to estimate the target $Y$ causes instability. Inspired by \cite{ref17}, we introduce an additional target network to stabilize the estimation of $Y$. Initially, we set up both a target and evaluation network. Throughout training, we focus on optimizing the evaluation network using our proposed loss functions, while keeping the target network parameters frozen. The target network is utilized exclusively to estimate the target $Y$ for the evaluation network. Following several iterations, we synchronize the parameters of the updated evaluation network with those of the target network.

The introduction of the target network serves to establish a reference point for the evaluation network, thereby stabilizing its performance during the training phase. For more details on the training process, refer to Algorithm~\ref{alg:training_process}. This approach ensures that the dynamic model undergoes stable and effective training, leading to improved performance and reliability in predicting future {room} temperatures.
\begin{algorithm}
    \caption{The procedure of training a dynamic model.}
    \begin{algorithmic}
        \REQUIRE{Training dataset $\boldsymbol{X}$; 
        Each training sample in $X$ consists of the form $\boldsymbol{[S_t, E_t, A_t, S_{t+\Delta}, E_{t+\Delta}]}$; 
        Learning rate $\boldsymbol{\alpha}$.}
        \ENSURE{Updated evaluation network parameters $\boldsymbol{\theta}$.}
        \STATE Randomly initialize target network $\theta_{Target}$ and evaluation network $\theta$, and freeze target network $\theta_{Target}$.
        \FOR{epoch $<$ EPOCH}
        \STATE Randomly sample N samples from $X$ as a batch.
        \FOR{every sample in BATCH}
        \STATE Compute $Loss_{STRB}$ (5), $Loss_{TDB}$ (13), where (10) is determined with $\theta_{\text{target}}$. Additionally, compute $Loss_{GC}$ (15), $Loss_{RC}$ (17), and $Loss_{recons}$ (18).
        \STATE Compute $Loss_{total}$ as in~\eqref{total_loss}.
        \STATE $\theta \gets \theta + \alpha\times\frac{1}{N}\times\frac{Loss_{total}}{\partial\theta}$.
        \ENDFOR
        \STATE Copy $\theta$ to target network: $\theta_{Target} \gets \theta$.
        \ENDFOR
    \end{algorithmic}
    \label{alg:training_process}
\end{algorithm}

\subsection{Control Strategy}\label{sec:control_strategy}

{Frequent or abrupt setpoint changes in HVAC systems increase energy consumption because sudden adjustments make the equipment work harder, often causing oscillations that lead to unnecessary energy waste. This occurs when control strategies focus only on short-term temperature targets, ignoring long-term stability.

The proposed long-term-aware setpoint selection strategy addresses this problem by balancing temperature alignment, long-term stability, and smooth transitions. It prioritizes long-term temperature stability using the predicted $S'_{\text{long}}$ and discourages abrupt setpoint changes. This reduces unnecessary setpoint fluctuations and indirectly lowers energy consumption while maintaining occupant comfort.}

{This setpoint selection strategy uses the dynamic model to} predict the future {room} temperatures $S'_{t+\Delta}$ given the current state, action and other uncontrollable factors. To determine the optimal setpoint, we utilize the dynamic model to predict future temperature for each setpoint in the action set $A$ based on {the} current state. For each setpoint, we calculate a score to decide which setpoint is optimal for {the} current situation. Notably, a previous study \cite{ref1} proposed to determine the setpoint by minimizing the error between the short-term temperature prediction $S'_{t+\Delta}$ and the expected temperature $T_{exp}$, which will easily encounter problems of setpoint oscillation. To tackle the problem, we take the long-term temperature prediction $S'_{t+\Delta}+D'_{t+\Delta}$ into consideration. To be more specific, the agent determines the optimal setpoint $A^*_t$ by maximizing the following scoring function:
\begin{equation}
    \label{temp_long}
    \begin{aligned}
        & S'_{long} = S'_{t+\Delta}+D'_{t+\Delta}
    \end{aligned}
\end{equation}

{This value shows the prediction of final stable room temperature that would be reached if a specific setpoint $a$ is maintained over time. After calculating $S_{\text{long}}'$ for each candidate, a scoring function is applied to evaluate the candidates based on current temperature alignment, long-term stability, and smooth transitions. The scoring function is defined as:}
\begin{equation}
    \label{temp_long_1}
    \begin{aligned}
        & Score = \left| S_t-T_{exp}\right|
        - \alpha\left| S'_{long}-T_{exp} \right| - \beta\left| S'_{long}-S_t \right|
    \end{aligned}
\end{equation}

{The scoring function balances HVAC control by considering current temperature alignment, long-term stability, and smooth transitions, which helps maintain occupant comfort, reduce oscillations, and improve energy efficiency. The hyperparameters $\alpha$ and $\beta$ are adjustable. Based on this evaluation, the candidate setpoint with the highest score is selected for discrete score-based optimization:
\begin{equation}
    \label{agent score}
    \begin{aligned}
        &A_{highest} = \mathop{\arg \max}_{a\in A}(Score)
    \end{aligned}
\end{equation}

where $Score$ is the score of candidate $a$. To determine the final setpoint, a confidence-based heuristic fusion combines $A_{\text{highest}}$ with the expert-defined setpoint $A_{\text{expert}}$ using a time-varying confidence $\alpha_t$, computed from the model's total prediction error \cite{ref1}:
\begin{equation}
    \label{alpha calculation}
    \begin{aligned}
\alpha_t = \beta \exp\Big(-\frac{loss_{Total}}{\sigma}\Big) + (1-\beta)\alpha_{t-1}
    \end{aligned}
\end{equation}
Here, $\beta$ and $\sigma$ are tunable parameters, and $\alpha_{t-1}$ smooths changes over time (initially $\alpha_0 = 0.5$). The final setpoint is then:
\begin{equation}
    \label{optimal setpoint}
    \begin{aligned}
A^*_t = \alpha_t \cdot A_{\text{highest}} + (1 - \alpha_t) \cdot A_{\text{expert}}
    \end{aligned}
\end{equation}

As shown in Algorithm~\ref{alg:setpoint_selection}, this strategy uses $S'_{\text{long}} = S'_{t+\Delta} + D'_{t+\Delta}$ instead of the short-term prediction $S'_{t+\Delta}$, focusing on long-term effects and avoiding short-term fluctuations. The confidence-based fusion ensures stable, safe, and adaptive control while reducing unnecessary setpoint fluctuations.
}

Although we have a well-trained model, the environment changes with seasons, rearranging the office, etc. Any changes {in} the environment may affect the accuracy of the model. Accordingly, the agent collects the new samples $[S_t, E_t, A_t, S_{t+\Delta}]$ during the operation phase, and utilizes them to adaptively update the model online. In this way, the agent can progressively adapt to the environment it is controlling as time advances.
\begin{algorithm}
    \caption{The procedure of setpoint selection.}
    \begin{algorithmic}
        \REQUIRE{The dynamic model $\boldsymbol{F}$ with the parameters $\boldsymbol\theta$; 
        The current state $\boldsymbol{S_t}$; 
        The current uncontrollable factor $\boldsymbol{E_t}$; 
        The confidence value $\boldsymbol{\alpha_t}$; 
        The expected temperature $\boldsymbol{T_{exp}}$; 
        The action set $\boldsymbol{A}$; 
        The expert's suggestion $\boldsymbol{A_{expert}}$.}
        \ENSURE{Optimal setpoint $\boldsymbol{A_{optimal}}$.}
        \STATE Define a empty score set $\boldsymbol{S}$.
        \FOR{each setpoint $a$ in $A$}
        \STATE $D'_{t+\Delta}, S'_{t+\Delta} = F(S_t, E_t, a;\theta)$
        \STATE $S'_{long} = D'_{t+\Delta} + S'_{t+\Delta}$
        \STATE $score = \left| S_t-T_{exp} \right| - \alpha\left| S'_{long}-T_{exp} \right| - \beta\left| S'_{long}-S_t \right|$.
        Append $score$ to $S$.
        \ENDFOR
        \STATE Choose the setpoint $A_{highest}$ that obtains the highest score in $S$.
        \STATE $A^*_t = \alpha_t \times A_{highest} + (1-\alpha_t) \times A_{expert}$.
    \end{algorithmic}
    \label{alg:setpoint_selection}
\end{algorithm}

\subsection{Pretrain and Transfer Learning}
In this section, the whole training process will be introduced. Overall, the training includes two main phases, namely the pretraining phase and the transfer learning phase. In the pretraining phase, we first construct a digital twin as the source domain based on our prior knowledge of the target building. Afterwards, we pretrain our model in the source domain dynamically. In the following section, we will elaborate on how the digital twin is tuned, how the pretraining works and last but not least, how we transfer the pre-trained model to target domain.

\subsubsection{Digital Twin Construction}
A digital twin serves as a virtual counterpart to real-world entities and processes. Leveraging a digital twin engine, we can simulate various technological scenarios and optimize critical decisions. In this study, we employ EnergyPlus as the carrier for physical entity representation, aiming to model the thermal environment within the building. 

\begin{figure}[!t]
    \centering{\includegraphics[width=\columnwidth]{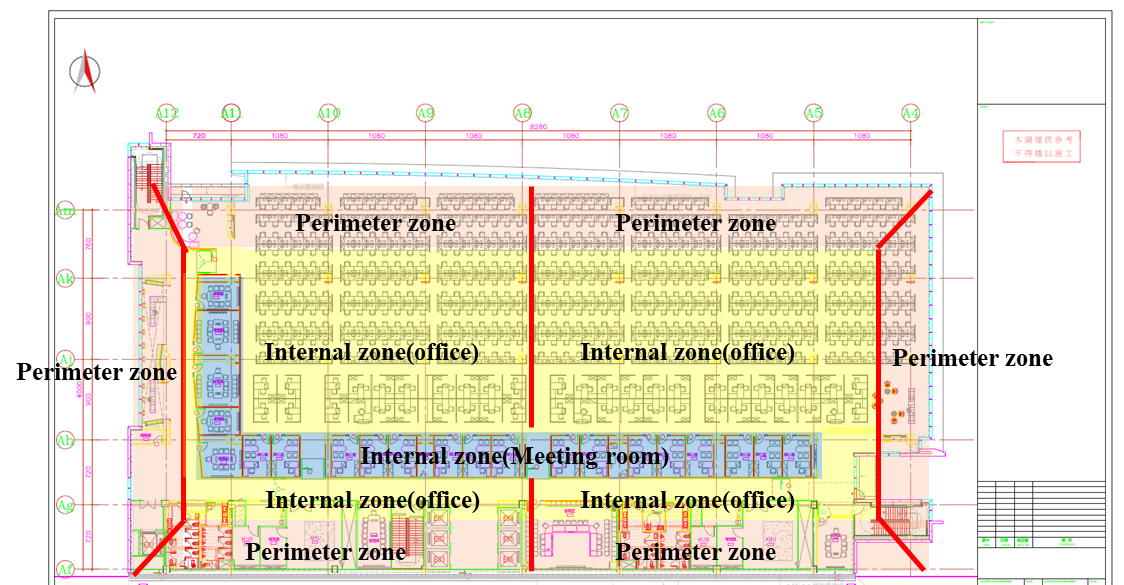}}
    \caption{Zone regionalization of the target building.}
    \label{figure:zone_regionalization}
\end{figure}

Initially, we construct the internal geometry of the target office building within the model. Given the inherent complexity of real-world internal conditions, we allow for some simplifications to focus on extracting thermal environment features. As illustrated in Fig.~\ref{figure:zone_regionalization}, our EnergyPlus model simulates both internal and perimeter zones, enabling separate analysis of weather and radiation effects on the building's periphery. Notably, the peripheral zone incorporates several Fan Coil Units (FCUs) to mitigate external weather impacts.

Subsequently, we meticulously set up the building envelope and internal design materials. Thermal parameters for the building envelope are derived from documentation provided by manufacturers, ensuring accuracy in simulation. Internal zones are demarcated using gypsum board, while the U-value of window glass is set to 0.84, based on profiles provided by window manufacturers \cite{ref21}.

{Furthermore, the Make-up Air Units (MAUs) supply fresh outdoor air to the internal zone Air Handling Unit (AHU), while Variable Air Volume (VAV) systems address localized cooling demands.  FCUs within the peripheral zone serve a similar purpose, catering to cooling needs influenced by external weather conditions.}

{
Occupancy schedules and internal loads follow typical weekday patterns during working hours and remain roughly the same over time, providing a consistent environment for our control agent. While information related to the target building remains relatively stable, weather plays a critical role in predicting indoor conditions. We incorporate local historical climate data from Taichung City, Taiwan, obtained from the Central Weather Administration. Together with geographical factors such as latitude, longitude, and sea level, we separate direct and diffuse horizontal irradiance to enhance the accuracy of our EnergyPlus simulations. Following rigorous checks and debugging iterations within EnergyPlus, we establish a robust building thermal environment model as the foundation for our digital twin.
}

 \begin{figure}[!t]
    \centering{\includegraphics[width=\columnwidth]{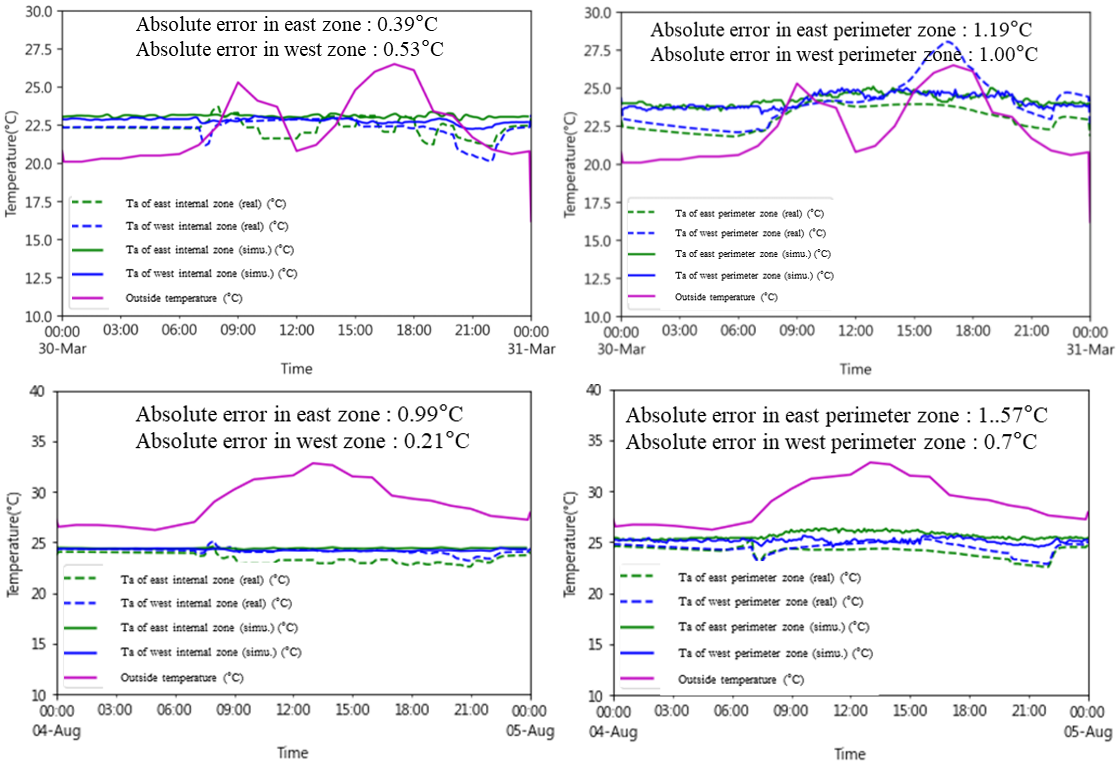}}
    \caption{The calibration result of the target building. The figures on the left-hand side: the results of internal zone. The figures on the right-hand side: the results of perimeter zone. Purple line: the outdoor temperature. Green/Blue line: temperature average of east/west zone. Dashed/Solid line: real/simulation data.}
    \label{figure:calibration_result}
\end{figure}

To validate the simulation results, we compare the {real-world} data with the simulated ones in both the internal and perimeter zones situated on the west and east sides of the target office space, as illustrated in {Fig.~\ref{figure:zone_regionalization}}. Across these four zones, we compare the air temperature data obtained from both real-world IoT sensors and the thermal environment model. The results, depicted in {Fig.~\ref{figure:calibration_result}}, reveal a calibration error ranging from 0.4 to 1.6$^\circ$C. Notably, the error tends to be higher in the perimeter zone, primarily due to the influence of outdoor air.

\subsubsection{Dynamic Pretraining}

\begin{figure}[!t]
    \centering{\includegraphics[width=\columnwidth]{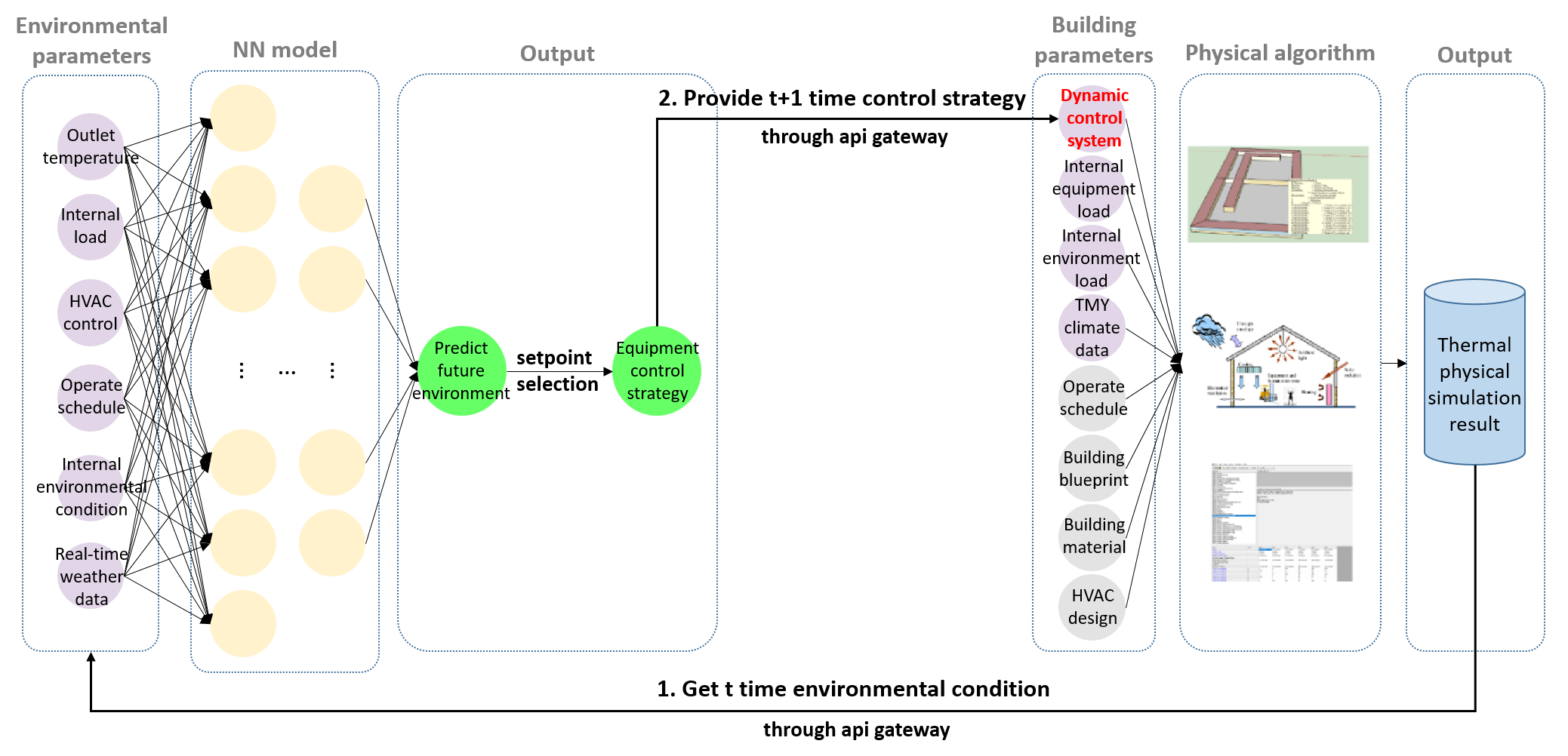}}
    \caption{The whole picture of pretrain scenario.}
    \label{figure:pretrain_scenario}
\end{figure}

Just like mentioned previously, the main difference between our proposed methods and other previous works utilizing EnergyPlus is that we train our model dynamically instead of statically with a fixed dataset. 

{This dynamic pretraining allows the agent to interact in real-time with the virtual building environment, continuously updating control strategies and observing thermal responses. This produces more relevant and diverse data for effective transfer learning.}

Specifically, we simulate the thermal dynamics whenever our agent makes new decisions. In our experiment setting, the agent decides {a} new setpoint every 18 minutes. Therefore, the setpoint schedule is updated in the 18-minutes interval. By doing so, the issue of outliers caused by unrealistic setpoint schedules can be avoid, since the schedules are constantly make by our own agent based on states at each time {step}. 

Fig.~\ref{figure:pretrain_scenario} presents the overall pipeline of pretraining scenario. Initially, the EnergyPlus model computes thermal physical simulation results based on the established building structure and input parameters, including the initial control schedule and local weather information. Subsequently, these simulation results are fed into the agent as input for the dynamic model to predict the next state. Based on this prediction, the agent generates a control action, which becomes the input parameters for the EnergyPlus model in the subsequent control period. Through the API connection, the EnergyPlus model and the agent interact to obtain necessary control schedules and environmental responses, mimicking real-world building dynamics.

With the online learning mechanism employed by the agent, the dynamic model continually updates and improves while controlling the HVAC system. This dynamic model pretraining occurs within the source domain, where the agent manipulates the virtual thermal environment to facilitate dynamic model learning. This pretraining strategy establishes a control target for the agent, mitigating outliers that could disrupt the dynamic model's learning process.

\subsubsection{Transfer Learning}

To mitigate biases between real and virtual environments, we make all dynamic model parameters trainable and fine-tune them using a small amount of real data from the target building. Despite challenges in obtaining rich and high-quality sensor data from reality, transfer learning enables knowledge acquired from the source domain to be transferred to the target domain, given the similarity between tasks in both domains. This ensures the model performs effectively in the target environment.

Additionally, to expedite system deployment, we implement transfers between real buildings. Leveraging an existing building with similar characteristics to the target new building, including size, layout, and HVAC system architecture, we transfer its stable system to the new target building. Following this, we fine-tune the system with a small amount of target building data to reconcile differences between the two buildings. This significantly reduces construction time, from over six months to within one week.

\section{Experimental Results}

The primary experimental site is located on the 7th floor of a tech company's office building in Hsinchu, Taiwan, comprising two open workspace areas designated as Room 701 and Room 702. To facilitate model-based transfer learning, both the source domain (a virtual environment constructed using EnergyPlus) and the target domain (Rooms 701 and 702) share a common set of 26 input variables. These variables include 5 VAV states ($S_t$), 20 uncontrollable environmental factors ($E_t$), and 1 AHU setpoint ($A_t$). Specifically, the uncontrollable factors encompass FCU temperatures, VAV and FCU switches, MAU temperature, outdoor temperature, rain status, and power consumption. A detailed description of each input sensor is provided in {Table~\ref{tab:input_description}}.

{In our case studies, the target room temperature is set to 25.25°C, which is an industrial requirement from our sponsor. Although actual sensor resolution may vary in practice, the datasets provided by the sponsor report temperatures with a precision of two decimal digits. This precise target allows quantitative evaluation of the control strategy, ensures consistent comparison against the reference, and demonstrates the accuracy of the proposed method. In practice, the target can be adjusted to match sensor resolution as preferred by users without affecting the performance of the control strategy.}

\begin{table}
    \caption{Sensor description}
    \label{tab:input_description}
    \setlength{\tabcolsep}{3pt}
    \resizebox{\columnwidth}{!}{%
        \begin{tabular}{p{98pt} c p{109pt}}
            \toprule
            \textbf{Sensor Type} & \textbf{Number} & \textbf{Description} \\  
            \midrule
            AHU setpoint ($A_t$) & 1 & Setpoint for the AHU outlet temperature (°C). \\
            VAV (room) temperature ($S_t$) & 5 & Return air temperature of VAVs (°C). \\
            FCU temperature ($E_t$) & 5 & Return air temperature of FCUs, located near windows (°C). \\
            VAV on/off ($E_t$) & 5 & Device status of VAVs (On=1, Off=0). \\
            FCU on/off ($E_t$) & 5 & Device status of FCUs (On=1, Off=0). \\
            Power supplied ($E_t$) & 2 & Sensors to measure the amount of supplied power (kW). \\
            MAU outlet temperature ($E_t$) & 1 & Sensor in the MAU to measure outlet air temperature in the air duct (°C). \\
            Outdoor temperature ($E_t$) & 1 & Sensor in the MAU to measure outdoor air temperature (°C). \\ 
            Weather condition ($E_t$) & 1 & Sensor outside the building to measure binary weather condition (Sunny=0, Rainy=1). \\
            \bottomrule
        \end{tabular}
    }
\end{table}

We possess two types of data: the first set comprises biased data collected through expert scheduling settings (illustrated in Fig.~\ref{figure:expert_control_rule}), while the second set is generated by our system during online control. In this online control scenario, setpoints may fluctuate depending on the agent's decisions, and the system records data at 18-minute intervals in the format $[S_t, E_t, A_t, S_{t+\Delta}]$. {Implementation details, including model architecture and training hyperparameters, are provided in Appendix~\ref{appendix:network}.}

\subsection{Evaluation Metric}

To ascertain the viability of the proposed HVAC control method and mitigate risks, access to a clean, diverse, and continuous historical dataset of building information is paramount. This dataset forms the bedrock for constructing an environmental simulation model, crucial for experimental endeavors. Regrettably, data availability for Room 701 and Room 702 is scant and incomplete. Consequently, we turned to the more comprehensive dataset from Room 301 to forge the environmental simulation model, addressing some gaps through linear interpolation.

Precisely, we employed 127 days of data gathered from Room 301 during the preceding online control period to establish the environmental model. This data was bifurcated: two-thirds constituted the training set, while the remaining one-third served as the test set. The model comprises fully-connected layers with VAV states, uncontrollable factors, and AHU setpoint as inputs, predicting VAV states for the subsequent time step. The mean prediction error of the environmental simulation model on the testing set approximates 0.0347°C. Gradient computation results suggest that the correlation between the input sensors and future {room} temperatures output is accurate, signifying the model's grasp of fundamental temperature change principles. Given the impracticality of conducting all desired experiments in a real office setting, the environmental simulation model serves as a surrogate, aiding in experimentation and analysis.

For evaluation purposes, we employ the following metrics:
\begin{equation}
    \label{DAE_metric}
    DAE = \frac{1}{n} \sum_{t=1}^n {\left| S_t-T_{exp} \right|}
\end{equation}
\begin{equation}
    \label{ME_metric}
    ME = \mathop{\arg\max}\limits_{i} DAE_i
\end{equation}
\begin{equation}
    \label{DARV_metric}
    DARV = \frac{1}{n-3} \sum_{t=3}^n {\left| (A_t-A_{t-1}) - (A_{t-1}-A_{t-2}) \right|}
\end{equation}

We calculate the daily average control error (DAE) to evaluate whether the indoor temperature aligns with the expected temperature. The one with the largest DAE during the control days is considered the maximum {control} error (ME). In {addition}, we compute the daily average regret value (DARV), which can be defined as the difference in control variable changes, to evaluate the smoothness of the setpoint variation. The larger the DARV, the greater the fluctuations in the setpoints. In summary, a good control method should be accurate and stable, that is, it should have low DAE and low DARV.

\begin{figure}
    \centering
    \subfigure[The DAE of different control methods.]{
            \begin{minipage}{6.75cm}
                \label{figure:compare_error_regretvalue:a}
                \includegraphics[width=\textwidth]{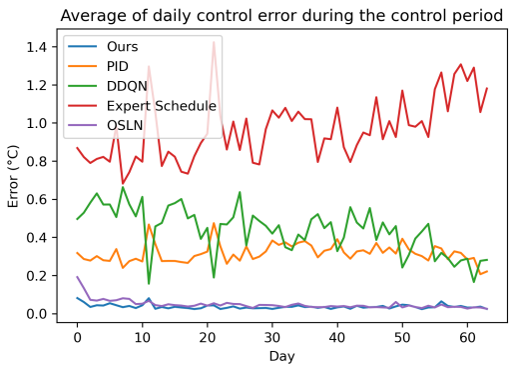}
            \end{minipage}
    }
    \subfigure[The DARV of different control methods.]{
            \begin{minipage}{6.75cm}
                \label{figure:compare_error_regretvalue:b}
                \includegraphics[width=\textwidth]{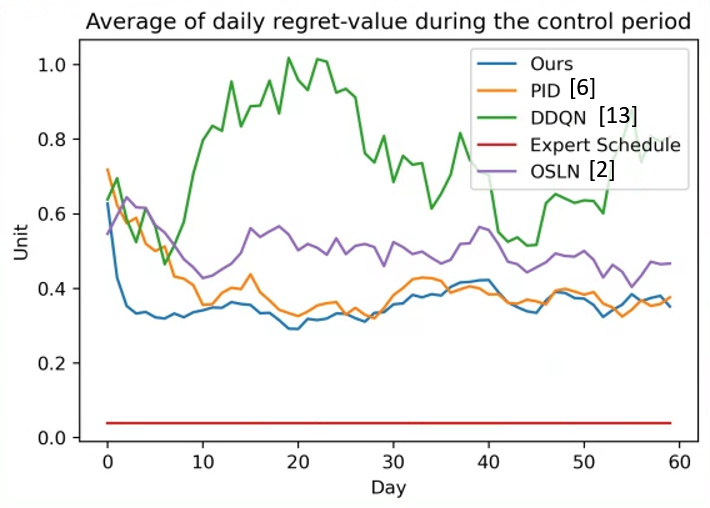}
            \end{minipage}
    }
    \label{figure:compare_error_regretvalue}
    \caption{Comparison of our method and other control methods.}
\end{figure}

\begin{figure*}
        \centering
        \includegraphics[width=\linewidth]{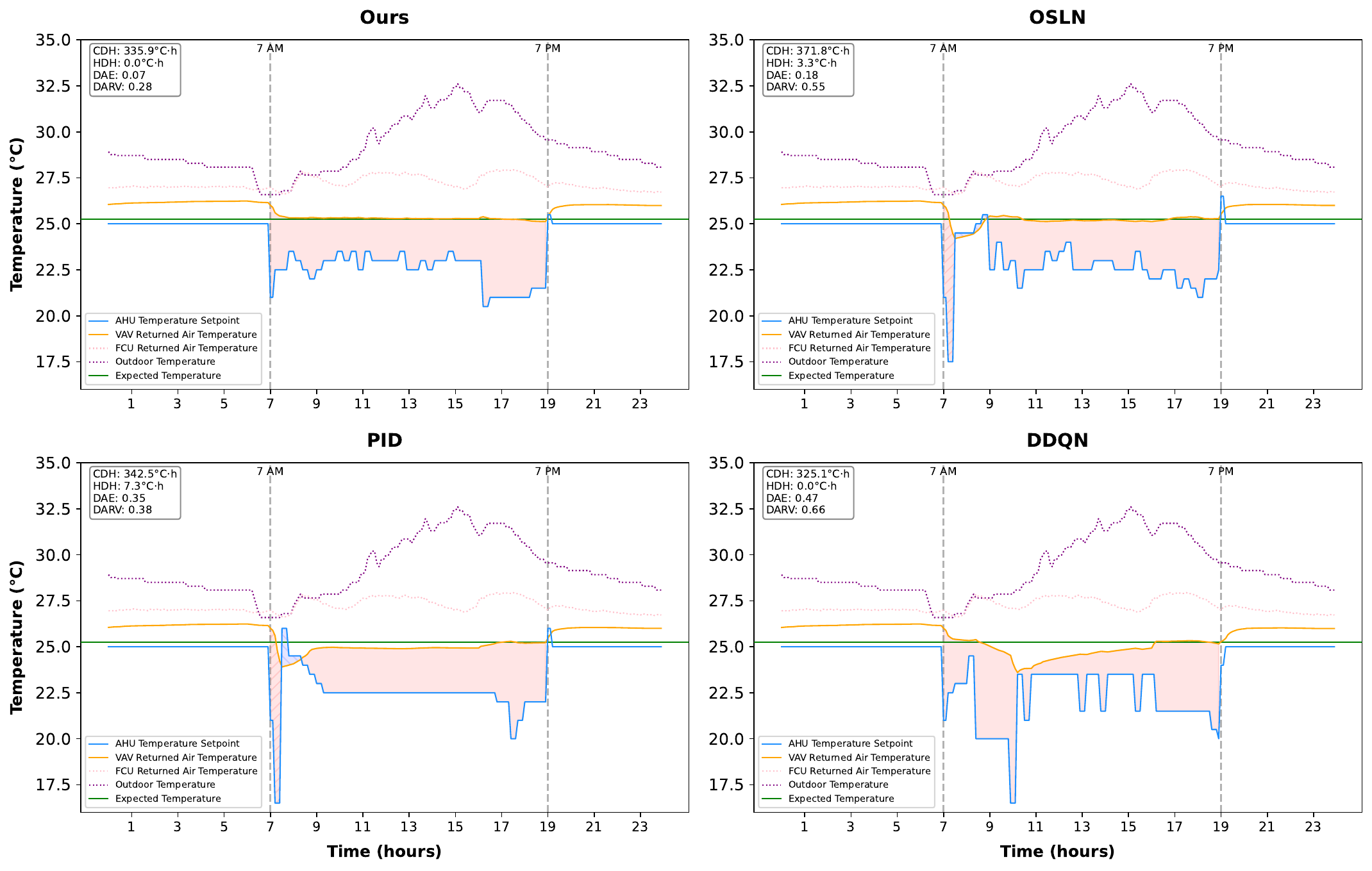}
        \caption{Comparison of four HVAC control methods during working hours. The red shaded area represents Cooling Degree Hours (CDH), and the blue shaded area represents Heating Degree Hours (HDH), illustrating the thermal gap between the AHU setpoint and the VAV returned air (room) temperature.}
        \label{fig:4method_energy_comparison}
\end{figure*}

\subsection{Comparison with other HVAC control methods}
Utilizing the environmental simulation model derived from historical data collected from Room 301, which was previously an experimental site for OSLN \cite{ref1} and accumulated extensive sensor data during agent control, we evaluated various HVAC control methods based on their performance metrics, specifically DAE and DARV. To facilitate comparison amidst significant oscillations in the curves, we applied a moving average smoothing technique to {Fig.~\ref{figure:compare_error_regretvalue:b}} (window size=5). Besides comparing with the original office setting (expert-defined schedule, as depicted in {Fig.~\ref{figure:expert_control_rule}}) and the traditional PID control method, we also include the advanced method \cite{ref1} and the RL-based approach \cite{ref23} for comparison. OSLN \cite{ref1} stands particularly relevant to our study as it represents a model-based technique that harnesses the benefits of RL-based methods, adapting over time to maintain indoor temperature close to the desired level efficiently. DDQN \cite{ref23}, on the other hand, is a hybrid RL method allowing a model-free agent to train within the system identification model.

The comparison depicted in {Fig.~\ref{figure:compare_error_regretvalue:a}} and {Fig.~\ref{figure:compare_error_regretvalue:b}} shows the superior performance and stability of our method. The expert-defined schedule, characterized by a temperature of 20.5$^\circ$C during working hours and 25$^\circ$C after work hours, yielded a control error exceeding 1$^\circ$C, indicating suboptimal performance. While the DDQN agent eventually stabilized over time due to continuous learning and updates, it exhibited heightened instability in the early stages, with consistently high regret values. Conversely, despite achieving reasonable control performance, PID control lacked environmental adaptability due to its limited understanding of the environment. OSLN demonstrated performance closest to ours, with an average control error of 0.17189$^\circ$C, while our method achieved a notably lower average control error of 0.03222$^\circ$C. Furthermore, our approach displayed lower control errors during the initial control stages. In Fig.~\ref{figure:compare_error_regretvalue:b}, our method exhibited a mean regret value of 0.37255, significantly lower than OSLN's mean regret value of 0.49868. This underscores the advantage of utilizing TDB to estimate the long-term impact of setpoints on the environment, enhancing the stability of setpoint trends.

{
\subsection{Energy Consumption Analysis}
To evaluate the proposed long-term-aware setpoint selection strategy, we analyze the energy consumption and performance of different HVAC methods. As noted in Section~\ref{sec:control_strategy}, abrupt setpoint changes increase energy waste and reduce comfort. By promoting smooth and stable setpoints, our method aims to reduce unnecessary energy waste while maintaining occupant comfort. To verify this, we compared four HVAC control strategies, including our method, OSLN, PID, and DDQN, over two weeks of continuous operation. The results are shown in Figure~\ref{fig:4method_energy_comparison}. We describe energy consumption in HVAC systems in two ways. Energy use refers to the total energy required to maintain comfort, while energy waste reflects additional energy spent due to setpoint fluctuations and oscillations. In this experiment, we quantify system deviation as the area between the AHU setpoint and the VAV return air temperature during working hours, with the area above the setpoint corresponding to Cooling Degree Hours (CDH) and the area below corresponding to Heating Degree Hours (HDH). Comfort is evaluated using the Daily Average Error (DAE), while fluctuations are assessed using the Daily Average Regret Values (DARV), reflecting how abrupt setpoint changes contribute to energy waste.

The results in Figure~\ref{fig:4method_energy_comparison} show clear differences in control behavior among the four HVAC methods. The OSLN and PID methods both show overcooling in the early hours, which increases CDH to 371.8°C·h and 342.5°C·h, respectively, because extra energy is needed to bring the temperature back to the target. Early oscillations in OSLN and moderate fluctuations later further increase energy waste, while PID keeps a relatively stable temperature around 22.5°C but still causes noticeable discomfort, with a DAE of 0.35. The DDQN method shows large oscillations throughout the day, slightly reducing CDH to 325.1°C·h but increasing energy waste and mechanical stress. Its high DAE of 0.47 and DARV of 0.66 indicate poor occupant comfort. In contrast, the proposed method keeps the temperature close to the target and achieves the lowest DAE of 0.07 with the smoothest control, shown by a DARV of 0.28. Although small fluctuations result in a CDH of 335.9°C·h, the proposed method reduces unnecessary energy use while maintaining high occupant comfort. By using long-term temperature prediction, it limits large oscillations and keeps temperature changes smooth, showing that energy-efficient control can be achieved even without directly including an energy term in the objective function.

}

\begin{figure*}[t]
    \centering{\includegraphics[width=\linewidth]{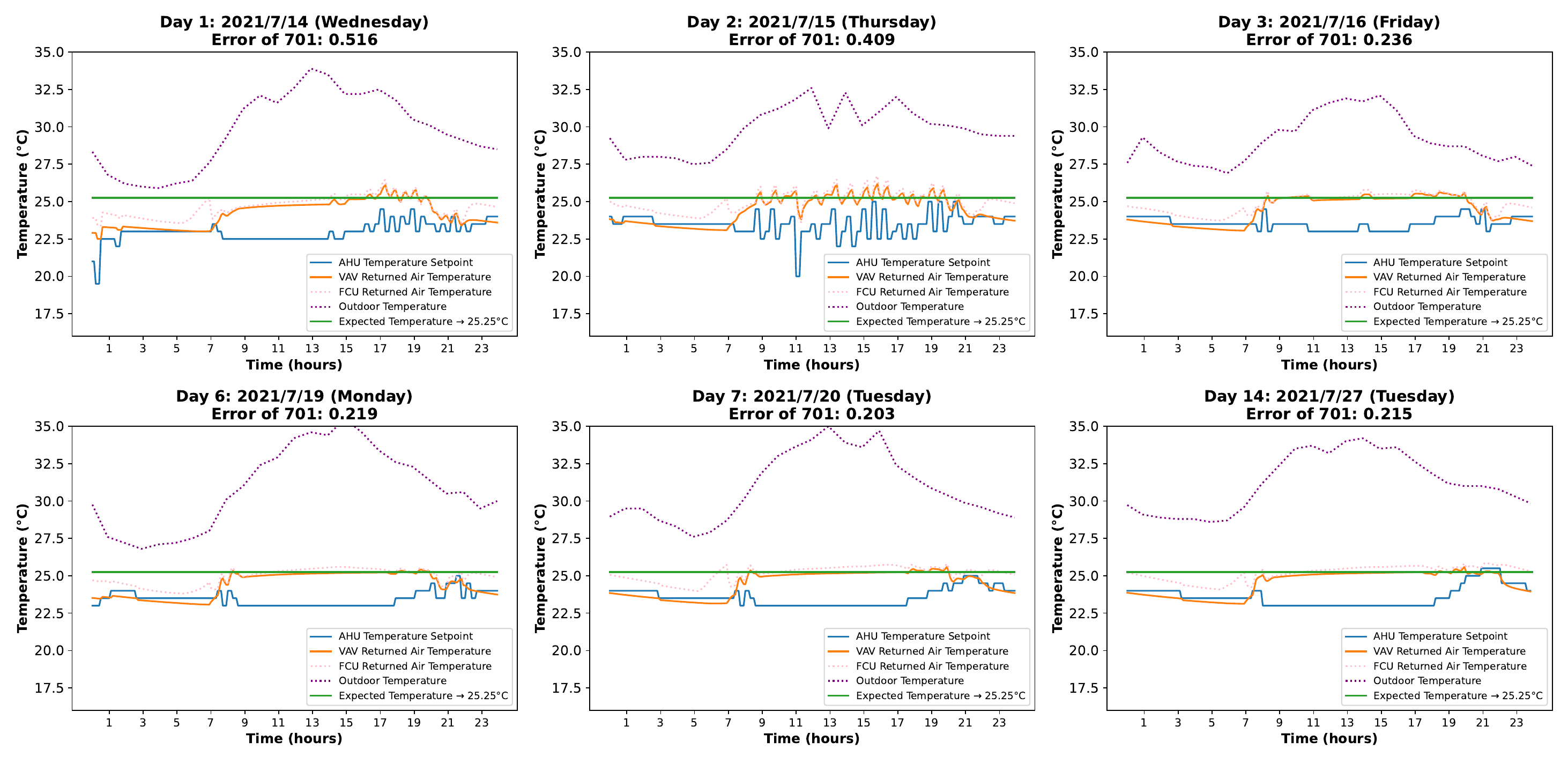}}
    \caption{The results of controlling the virtual environment of Room 701. (From left to right, top to bottom, these are the first, second, third, sixth, seventh, and fourteenth days of the control.) Green line: the expected temperature. Blue line: the setpoint. Orange line: the room temperature. Pink line: the temperature near the window. Purple line: the outdoor temperature. The error is calculated as the difference between the expected temperature and the room temperature, and is only considered during working hours (from 7am to 7pm).}
    \label{figure:701_simulated_control_EnergyPlus}
\end{figure*}

\subsection{Transfer learning}
\subsubsection{Dynamic pretraining in the source domain}
We utilized EnergyPlus to construct virtual environments for {Rooms} 701 and 702, allowing our agent to control these environments. Operating on an 18-minute control cycle, resembling the real-world settings, the target temperature was set at 25.25$^\circ$C. The simulation results, depicted in 
Fig.~\ref{figure:701_simulated_control_EnergyPlus}, {show the outcomes of controlling the environment of room 701, with behaviours that align closely with real-world dynamics. {The model begins with randomly initialized weights, and the agent selects setpoints within a predefined range. As the agent interacts with the virtual environment and updates its actions through online learning, it gradually adapts to the thermal dynamics of the room}. Notably setpoints tend to undergo pronounced adjusments when the working hours about to start and end, especially between 5pm and 9 pm. Room 702 followed a similar trend, undergoing the same training process and exhibiting similar control patterns throughout the simulation. {Weather data and building dynamics are incorporated to ensure the model experiences realistic scenarios, including extreme or rare conditions that would be unsafe to test in a real building.}}

By training the dynamic model through controlling virtual environments, it acquires insights into temperature variations. These learned characteristics serve as initial parameters for the target domain model, effectively transferring knowledge of temperature dynamics. Subsequently, the model undergoes fine-tuning using a small quantity of real sensor data, ensuring alignment with real-world conditions.

\subsubsection{Comparison with other transfer learning methods}

We compared our proposed method and two established HVAC control methodologies: FDTDM \cite{ref7}, a model-based transfer learning approach, and DAF \cite{ref12}, a domain adaptation framework. Model-based transfer learning methods are prevalent in the field of HVAC control, with FDTDM being a notable example. Following pretraining in the source domain, FDTDM offers two fine-tuning options: freezing either the initial layer (referred to as the "head") or the final layer (referred to as the "tail"). Conversely, DAF is a feature-based transfer learning technique that maps domain-invariant features to a shared space, facilitating knowledge transfer from the source to the target domain.

{In our implementation of FDTDM and DAF, we utilized EnergyPlus-generated simulation data available in July 2021 as the source domain, while the target domain comprised 5 days of biased data collected from the real-world office environment. Similarly, our proposed method leveraged dynamic pretraining, allowing the agent to control the source domain using data available in July 2021. During fine-tuning in the target domain, we employed the same 5 days of biased real-world office data}

For evaluation, we employed real office data collected during the system's online control phase as the test dataset. This dataset encompassed {90 days categorized as summer (from July to October) and 70 days categorized as winter (from December to April)}. The results, presented in {Table~\ref{tab:compare_TL_method}}, demonstrate that our transfer learning approach enables dynamic learning within the source domain, employing control conditions identical to those of the target domain. Compared to alternative transfer learning methods, our approach facilitates learning from samples more akin to the target domain, consequently outperforming both FDTDM and DAF.

\begin{table}[t]
    \centering
    \caption{Comparison of our method and other TL methods}
    \label{tab:compare_TL_method}
    \resizebox{\columnwidth}{!}{%
        \begin{tabular}{lcccccc}
            \toprule
            \multirow{3}*{Method} & \multicolumn{3}{c}{Room 701} & \multicolumn{3}{c}{Room 702} \\ 
            & Summer & Winter & Avg. & Summer & Winter & Avg. \\
            \midrule
            \makecell[l]{FATDM\\(head)\cite{ref7}} & 0.451    & 0.952    & 0.701    & 0.570    & 0.716    & 0.642 \\
            \makecell[l]{FATDM\\(tail)\cite{ref7}} & 0.517    & 0.962    & 0.740    & 0.488    & 0.622    & 0.555 \\
            DAF\cite{ref12}   & 0.598    & 0.605    & 0.602    & 0.545    & 0.663    & 0.604 \\
            Ours  & \textbf{0.078}    & \textbf{0.202}   & \textbf{0.140}    &\textbf{0.056}    & \textbf{0.298}   & \textbf{0.177} \\
            \bottomrule
        \end{tabular}
    }
\end{table}
 
\subsection{Ablation study}
\subsubsection{The agent}

{We explore the effect of tuning various hyperparameters, with results shown in {Table~\ref{tab:hyperparameter}}. In Algorithm~\ref{alg:setpoint_selection}, $\alpha$ and $\beta$ respectively indicate whether the estimated long-term average temperature aligns with the user's target and whether it causes significant changes in the current environmental state under the $A_t$ setting. A higher $\gamma$ value indicates that recent temperatures have a greater influence, meaning that the long-term average temperature will be closer to the most recent temperatures, and vice versa. In {Table~\ref{tab:hyperparameter}}, we observe that reducing $\alpha$ (e.g., to 0.5 or 0.75) leads to a sharp increase in control error (ME $= 0.9175$), while lowering $\beta$ results in a clear increase in regret values. This indicate that a large gap between the present and future states causes drastic changes in setpoints. Conversely, setting $\beta$ too high (e.g., 1.0) increases ME, suggesting a trade-off between stability and accuracy.} 

{Regarding $\gamma$, increasing it beyond 0.25 results in a slight improvement in ME and a notable decrease in DARV, indicating better long-term stability. Although DAE experiences a slight increase, this trade-off indicates that $\gamma = 0.5$ offers the most effective balance between control accuracy and overall system robustness. This outcome aligns with the characteristics observed in OSLN~\cite{ref1}, which emphasizes short-term optimization without compromising long-term reliability. In conclusion, selecting $\alpha = 1.0$, $\beta = 0.9$, and $\gamma = 0.5$ provides the optimal trade-off between accuracy and stability.}

\begin{table}[t]
    \centering
    \caption{Ablation studies}
    \label{tab:hyperparameter}
    \resizebox{\columnwidth}{!}{%
        \begin{tabular}{cccccc}
            \toprule
            \textbf{$\alpha$} & \textbf{$\beta$} & \textbf{$\gamma$} & \textbf{ME} & \textbf{Avg. DAE} & \textbf{Avg. DARV} \\
            \midrule
            0.5 & 0.9 & 0.25 & 0.9175 & 0.3433 & 0.3151 \\
            0.75 & 0.9 & 0.25 & 0.9175 & 0.2966 & 0.3058 \\
            1.0 & 0.9 & 0.25 & 0.2522 & 0.0566 & 0.3851 \\
            1.1 & 0.9 & 0.25 & 0.2522 & 0.0558 & 0.3954 \\
            1.5 & 0.9 & 0.25 & 0.2521 & 0.0531 & 0.3939 \\
            \midrule
            1.0 & 0.5 & 0.25 & 0.2556 & 0.0544 & 0.4058 \\
            1.0 & 0.75 & 0.25 & 0.2526 & 0.0573 & 0.3880 \\
            1.0 & 0.9 & 0.25 & 0.2522 & 0.0566 & 0.3851 \\
            1.0 & 1.0 & 0.25 & 1.2887 & 0.1404 & 0.3285 \\
            1.0 & 1.1 & 0.25 & 0.9175 & 0.2770 & 0.3193 \\
            \midrule
            1.0 & 0.9 & 0.10 & 0.2360 & 0.0546 & 0.4041 \\
            1.0 & 0.9 & 0.25 & 0.2522 & 0.0566 & 0.3851 \\
            \textbf{1.0} & \textbf{0.9} & \textbf{0.50} & \textbf{0.2322} & \textbf{0.0613} & \textbf{0.3562} \\
            1.0 & 0.9 & 0.75 & 0.2343 & 0.0774 & 0.2650 \\
            \bottomrule
        \end{tabular}
    }
\end{table}

\begin{table}[t]
    \centering
    \caption{Ablation studies}
    \label{tab:ablation_study}
    \resizebox{\columnwidth}{!}{%
        \begin{tabular}{lccc}
            \toprule
            \textbf{Method} & \textbf{ME} & \textbf{Avg. DAE} & \textbf{Avg. DARV} \\ \midrule
            w/o physical constraint loss & 0.2656          & 0.0961          & \textbf{0.3463} \\
            w/o TDB             & 0.2448          & \textbf{0.0558} & 0.4110 \\
            w/o online learning & 0.6778          & 0.2978          & 0.9602 \\
            w/o confidence value & 0.3691         & 0.0667          & 0.3516 \\
            Ours        & \textbf{0.2322}         & 0.0613          & 0.3562 \\
            \bottomrule
        \end{tabular}
    }
\end{table}

{Additionally, to analyze the impact of each mechanism in our method, we conducted ablation studies using the environmental simulation model, which serves as a proxy for a real-world environment. We established five experimental settings: our final proposed agent, the exclusion of the both physical constraints loss: gradient constraint loss~\eqref{gradient_constraint_loss} and residual constraint loss~\eqref{residual_constraint_loss} during model training, the removal of TDB~\eqref{difference_loss} (employing the minimum error between short-term temperature predictions and expected temperature to select actions), the elimination of the online learning mechanism, and the absence of confidence value consideration (directly outputting the highest-scoring setpoint). In all cases, these methods were initially trained using an equal amount of biased data before deployment into the environmental simulation model for control.}

{As shown in Table~\ref{tab:ablation_study}, the impact of removing each mechanism on control performance is evident. Eliminating the both physical constraints (the gradient constraint \eqref{gradient_constraint_loss} and the residual constraint \eqref{residual_constraint_loss}) results in increased model prediction errors under biased data influence, underscoring the role of embedding physical rules in mitigating training data bias. 

{Removing TDB forces the system to rely on short-term predictions, which results in higher DARV and more frequent or larger setpoint changes, showing that the agent cannot maintain long-term stability without this guidance. Including TDB allows the model to balance short-term accuracy with long-term stability. It guides the agent select setpoints that avoid abrupt changes and provides smoother temperature transitions, reduced fluctuations, and maintained occupant comfort.}

Omitting the online learning mechanism and confidence value assessment leads to a lack of continuous environmental adaptability and risk management, resulting in a notable performance decline. Conversely, our final proposed agent outperforms across metrics, exhibiting the best ME during the control period and achieving a balanced trade-off between DAE and DARV.}

\subsubsection{Transfer learning}

To underscore the advantages of the proposed dynamic transferred model, we comapare it with two other models: one trained from scratch, which has not undergone pretraining on the source domain, and a static transferred model, which is pre-trained in a fixed manner. The training data for the model trained from scratch is identical to the target domain data, comprising biased data totaling approximately 720 samples, which is used for fine-tuning the transferred model. Hence, to be precise, the sole difference between the model trained from scratch and the transferred model lies in whether they have undergone pretraining on the source domain or not.

\begin{figure*}[t!]
    \centering
    \includegraphics[width=1\linewidth]{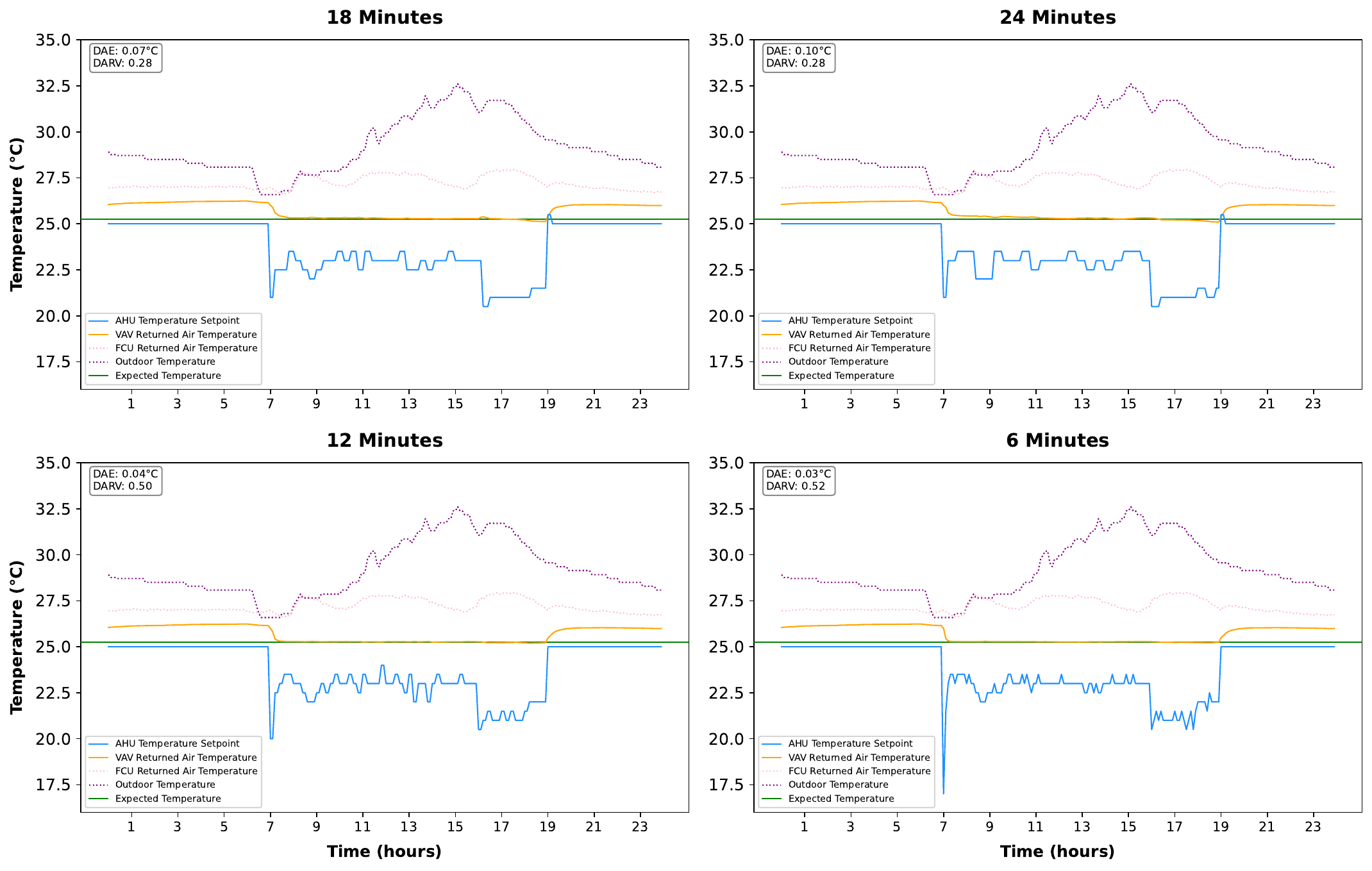}
    \caption{Comparison of temperature control performance for 6, 12, 18, and 24-minute control cycles. The 18-minute cycle achieves the best balance between stability and accuracy.}
    \label{fig:app_control_cycles}
\end{figure*}

\begin{table}[t]
    \centering
    \caption{Ablation studies}
    \label{tab:ablation_study_TL}
    \resizebox{\columnwidth}{!}{%
        \begin{tabular}{lcccccc}
            \toprule
            \multirow{3}*{Method} & \multicolumn{3}{c}{Room 701} & \multicolumn{3}{c}{Room 702} \\ 
            & Summer & Winter & Avg. & Summer & Winter & Avg. \\
            \midrule
            w/o TL    & 0.451    & 0.359    & 0.405    & 0.259    & 0.379    & 0.319 \\
            static    & 0.215    & 0.396    & 0.306    & 0.204    & 0.346    & 0.275 \\
            Ours  & \textbf{0.078}    & \textbf{0.202}   & \textbf{0.140}  
                &\textbf{0.056}    & \textbf{0.298}   & \textbf{0.177} \\
            \bottomrule
        \end{tabular}
    }
\end{table}

As indicated in Table~\ref{tab:ablation_study_TL}, it is evident that the model trained from scratch, without leveraging the source domain, exhibits significantly higher errors compared to the transferred models, underscoring the necessity of transfer learning. Furthermore, upon comparing the static transferred model with our dynamic transferred model, it becomes apparent that our method yields substantially lower errors, thereby highlighting the effectiveness of the proposed dynamic pretraining approach.

\subsubsection{Control Cycle Duration}\label{subsubsec:control_cycle_duration}

{To determine the optimal control cycle for short-term temperature prediction, we compared 6, 12, 18, and 24-minute cycles over two weeks of continuous control, as shown in Figure~\ref{fig:app_control_cycles}. Short cycles (6 or 12 minutes) caused frequent and abrupt AHU adjustments, which led to large temperature oscillations and reduced system stability. Although the prediction error is lower, the system does not operate smoothly and may cause mechanical wear and energy waste.

Long cycles (24 minutes) reduced the frequency of AHU adjustments, providing smoother operation, but the system response became slower and prediction error increased. In our experiments, the 24-minute cycle had a similar DARV to the 18-minute cycle (0.28), but the DAE increased to 0.10, indicating lower accuracy. The 18-minute cycle provided the best balance between stability and accuracy, with a DARV of 0.28 and a DAE of 0.07, avoiding both the oscillations of short cycles and the slow response of long cycles. This choice is also consistent with HVAC practice, where room temperature typically converges within approximately 18 minutes after an AHU setpoint change, and aligns with prior studies on short-term HVAC control \cite{ref1}.
}

\begin{table}[t]
    \centering
    \caption{Comparison of Fully-Connected ED Model and Time-Series Models}
    \label{tab:time_dependent_model_comparison_time}
    \resizebox{\columnwidth}{!}{%
    \begin{tabular}{lccccc}
        \toprule
        \textbf{Method} & \textbf{Sequence Length} & \textbf{Num Layers} & \textbf{DAE} & \textbf{DARV} & \textbf{Time} \\ 
        \midrule
        \multicolumn{6}{c}{{\textit{Fully-Connected Encoder-Decoder Model}}} \\ 
        \midrule
        Ours      & - & - & \textbf{0.03222} & 0.37255 & \textbf{0.012} \\
        \midrule
        \multicolumn{6}{c}{{\textit{Time-Series Models}}} \\ 
        \midrule
        \multirow{4}{*}{LSTM} 
                  & 2 & 2 & 0.6182 & \underline{0.2389} & \underline{0.028} \\
                  & 3 & 2 & 0.8349 & 0.2453 & 0.034 \\
                  & 3 & 4 & 1.1604 & \textbf{0.1630} & 0.052 \\
                  & 6 & 4 & 0.9492 & 0.2893 & 0.092 \\
        \midrule
        \multirow{4}{*}{Transformer} 
                  & 2 & 2 & 1.0243 & 0.3612 & 0.108 \\
                  & 3 & 2 & 0.8984 & 0.3362 & 0.136 \\
                  & 3 & 4 & 0.9378 & 0.3641 & 0.258 \\
                  & 6 & 4 & \underline{0.6069} & 0.4039 & 0.621 \\
        \bottomrule
    \end{tabular}
    }
\end{table}

\subsubsection{Comparison with Time-Series Models}

{
We conducted an ablation study to compare the performance of fully-connected encoder-decoder (ED) models with time-series models, including LSTM and Transformer, for HVAC control. In these experiments, unidirectional LSTM layers and encoder-only Transformer layers were used to process recent sequences of state-action pairs and predict the next state, with a fully-connected layer producing the final output. 

As shown in Table~\ref{tab:time_dependent_model_comparison_time}, time-series models are able to capture temporal dependencies, resulting in slightly lower or comparable DARV values in some configurations. For instance, the LSTM with three layers and a sequence length of four achieves the lowest DARV of 0.1630. However, these models consistently exhibit higher DAE, indicating poorer short-term prediction accuracy. Transformer models show similar trends and additionally incur substantially higher inference time, which can limit their use in real-time control applications.

In contrast, the fully-connected ED model with the Temperature Difference Branch (TDB) achieves both low DAE and competitive DARV, while requiring minimal computational resources. This design captures the differences between long-term and short-term temperature patterns without relying on historical sequences, providing stable control and smooth temperature transitions. These results highlight that, while time-series models can model temporal dynamics, our fully-connected ED design achieves a better balance between short-term prediction accuracy, long-term stability, and computational efficiency. This ablation study supports the choice of the ED model as the main architecture for our HVAC controller.
}

\subsection{Applications in real environment}
We utilized a pre-trained agent in the EnergyPlus virtual environment, fine-tuning it with a small amount of real-world environment data that exhibited some bias. Subsequently, we deployed the agent into the actual environments of Room 701 and Room 702.

\begin{figure}
    \centering{\includegraphics[width=\columnwidth]{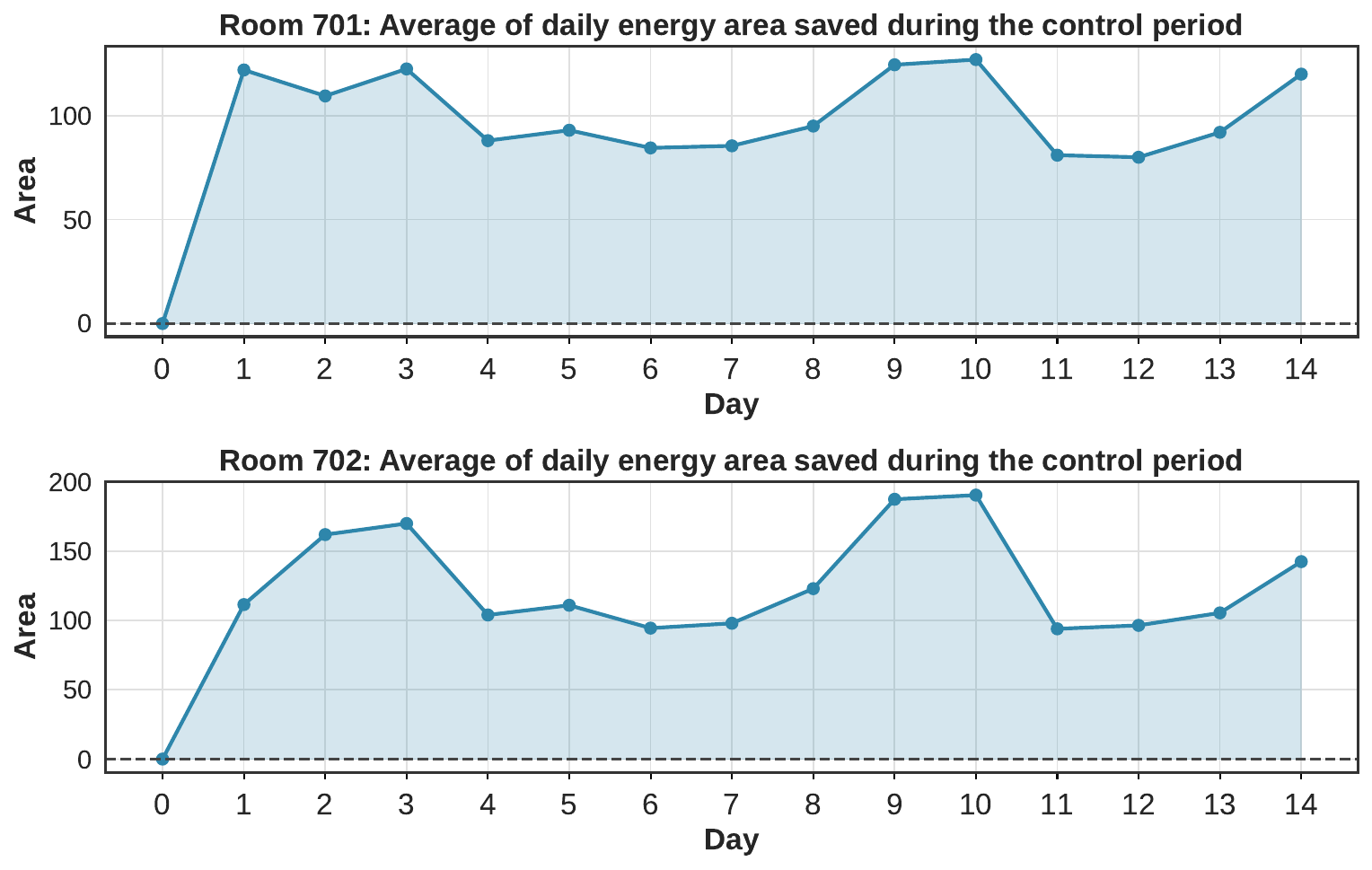}}
    \caption{The area between the historical setpoint curves and the setpoint curves controlled by the agent. The upper figure shows the experiment from the 1st day to the 14th day in Room 701, while the lower one pertains to Room 702.}
    \label{figure:energy_saved}
\end{figure}

In Fig.~\ref{figure:energy_saved}, we computed the area between the historical full-day setpoint schedule curves for these two offices (both set to 20.5$^\circ$C in the past) and the full-day setpoint schedule curves controlled by the agent. With the time unit set at 18 minutes (control cycle), a larger area indicates that the setpoints decided by the agent were higher than the original settings, resulting in greater energy savings. Our control method demonstrates significant energy efficiency benefits for the office environment.

\begin{figure}
    \centering{\includegraphics[width=8.8cm]{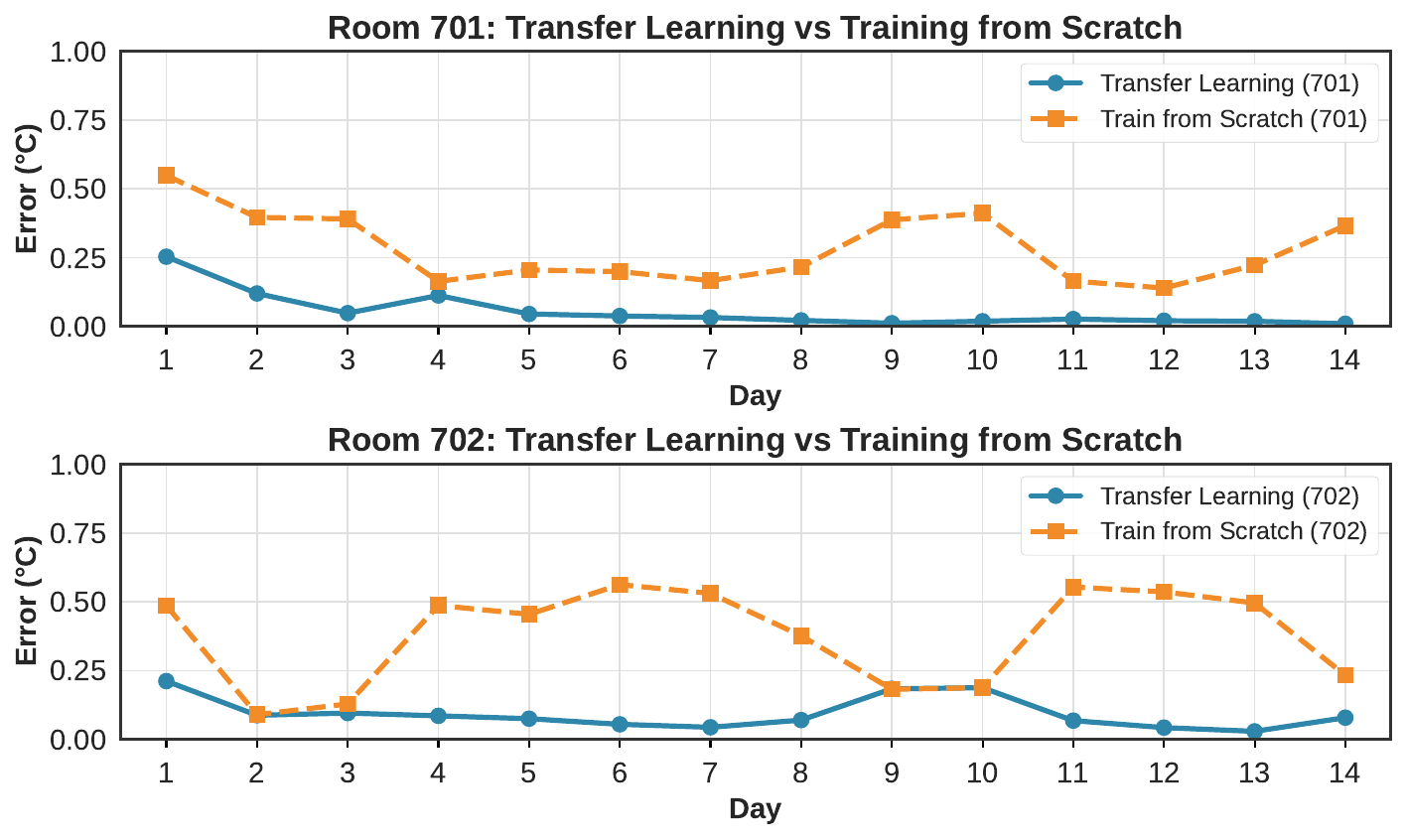}}
    \caption{Comparison of transferred models and models trained from scratch. The figure shows the DAE (difference between room temperature and expected temperature) for each model from the 1st day to the 14th day of the experiment. The upper figure shows the experiment on Room 701, and the lower one shows on Room 702.}
    \label{figure:compare_online_model}
\end{figure}

Additionally, we compared the models trained from scratch with the dynamic transferred models. {Both sets of experiments were conducted using real office data collected during the summer season, ensuring comparable environmental conditions}. As illustrated in Fig.~\ref{figure:compare_online_model}, the experimental results show that the dynamic transferred models exhibit lower control errors in the early stages of control and overall, they appear to be more stable and accurate compared to the models trained from scratch. This demonstrates that under the same target domain data (utilizing five-day bias data, totaling 1200 samples in our experiment), dynamic transferred models achieve superior control performance. The advantage of transfer learning lies in reducing the model's reliance on target domain data, leading to improved performance compared to models trained from scratch. Moreover, it significantly reduces the time and cost of collecting target domain data, facilitating practical applications and rapid expansion.

\begin{figure}
    \centering{\includegraphics[width=\columnwidth]{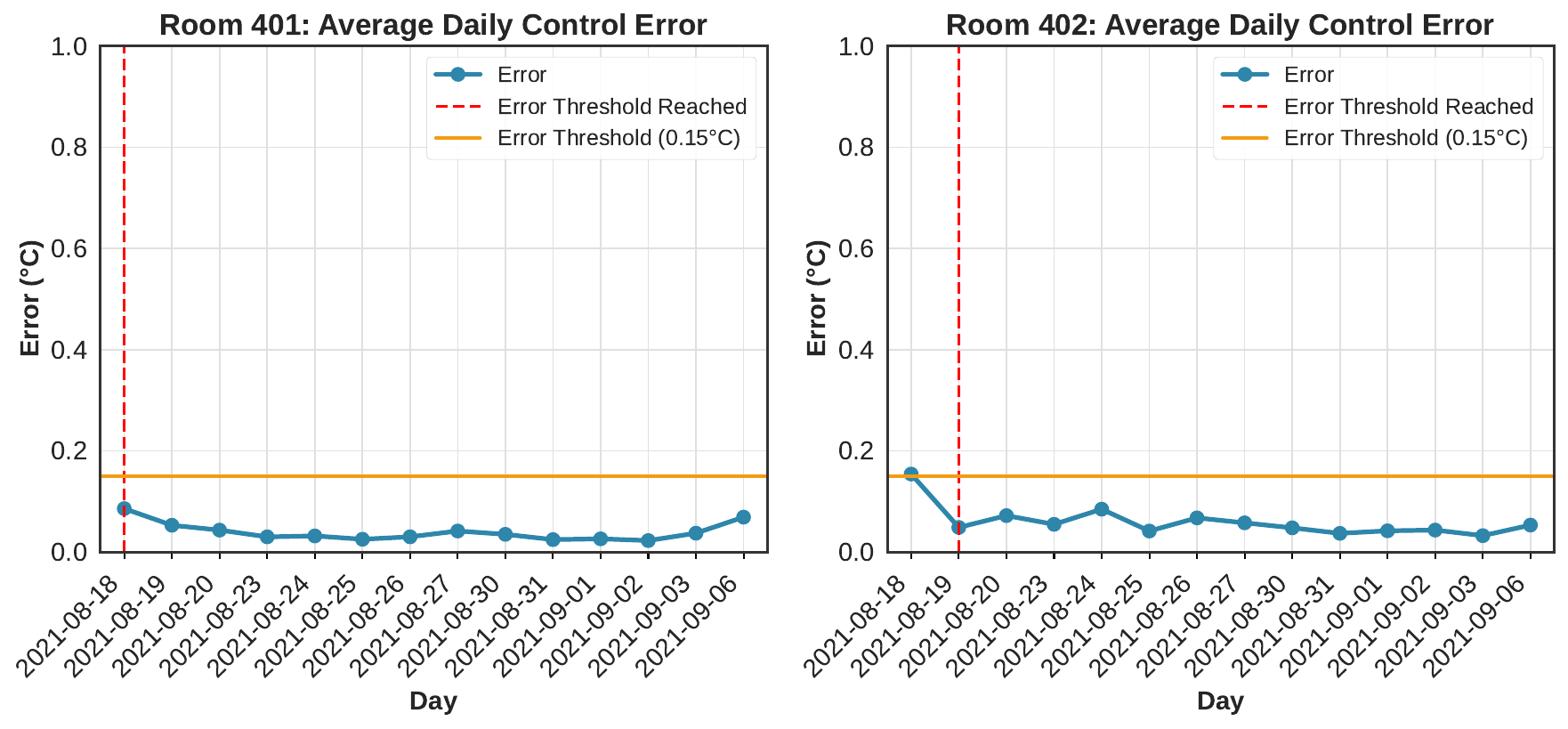}}
    \caption{The transfer between real-world offices. Room 301 serves as the common source domain, while Room 401 and Room 402 are the target domains. The experiment started on August 18th, 2021. The figure illustrates the performance using DAE.}
    \label{figure:real_world_TL}
\end{figure}

It's important to note that the experimental sites share similar characteristics, all being office buildings with access to similar sensors. To expedite the stable deployment of agents in new offices, we employed agents from other existing offices where they were already operational and stable. By transferring these agents to the new office, we saved time required for creating a virtual office, thus achieving a faster deployment process. In this experiment, we considered Room 301, where the agent had been operational since 2018, as the source domain. Room 401 and Room 402 were treated as the target domains. Following the same transfer learning method we proposed, we transferred the Room 301 model to be used in Room 401 and Room 402. The results, as shown in Fig.~\ref{figure:real_world_TL}, demonstrate that agents operating in Room 401 and Room 402 were both able to reduce control errors to below 0.15$^\circ$C within one week.

\section{Conclusion}
{We introduce an adaptive model-based transfer learning approach that integrates physics rules and a long-term-aware setpoint selection strategy to achieve precise, stable, and energy-efficient HVAC control. Compared with OSLN, PID, and DDQN, our method consistently delivers superior occupant comfort, stable control, and smooth setpoint transitions while reducing abrupt changes that cause energy waste. The adaptive transfer learning framework enables dynamic pretraining in the source domain and fine-tuning with limited target domain data, allowing faster deployment and improved control performance in real-world office environments with minimal reliance on extensive target-specific knowledge. Ablation studies demonstrate that physical constraints, the Temperature Difference Branch (TDB), online learning, and confidence value assessment are essential for achieving stable, accurate, and adaptable control. Real-world experiments show that the proposed method not only maintains occupant comfort but also significantly improves energy efficiency compared with historical schedules. The experimental outcomes underscore the effectiveness of our proposed HVAC control methodology and transfer learning approach, affirming their practical relevance and performance in real-world scenarios.}

\bibliographystyle{IEEEtran}
\bibliography{bibliography}

\newpage

\begin{appendices}
\section{Hyperparameters for Model and Training Settings}
{\label{appendix:network}}
We provide a table to summarize the hyperparameters used in our work, including both model and training settings
\begin{table}[!h]
    \centering
    \label{tab:hyperparameters}
    \resizebox{\columnwidth}{!}{%
    \begin{tabular}{lc}
        \toprule
        \textbf{Hyperparameter} & \textbf{Value} \\
        \midrule
        \multicolumn{2}{c}{\textit{Encoder Hyperparameters}} \\
        \midrule
        Layers & 2 linear layers per state/action branch \\
        Hidden units & 32 units per layer \\
        Activation function & LeakyReLU (0.2) per layer \\
        \midrule
        \multicolumn{2}{c}{\textit{Predictor Hyperparameters}} \\
        \midrule
        Shared layers & 3 linear layers \\
        Shared hidden units & 64 units per layer \\
        TDB and STRB branch layers & 2 linear layers per TDB/STRB branch \\
        Branch hidden units & 64 and 32 units \\
        Activation function & LeakyReLU (0.2) per layer \\
        \midrule
        \multicolumn{2}{c}{\textit{Decoder Hyperparameters}} \\
        \midrule
        Layers & 1 linear layer \\
        Hidden units & 32 units \\
        \midrule
        \multicolumn{2}{c}{\textit{Training Settings}} \\
        \midrule
        Replay buffer size & 42 × 240 samples \\
        Target network update interval & Every 100 steps \\
        Optimizer & Adam \\
        Learning rate & 0.001 \\
        Batch size & 1000 \\
        \bottomrule
    \end{tabular}
    }
\end{table}

\end{appendices}

\end{document}